\documentclass[a4paper,11pt]{article}
\pdfoutput=1 % if your are submitting a pdflatex (i.e. if you have
\usepackage{jcappub} % for details on the use of the package, please
\usepackage{graphicx}% Include figure files
\usepackage{dcolumn}% Align table columns on decimal point
\usepackage{bm}
\usepackage{amsmath,amsfonts,amssymb}
\usepackage{graphicx,epsfig}
\usepackage{color}
\usepackage{captcont}
\usepackage{slashed}
\usepackage{epstopdf}
\usepackage{mathtools}
\usepackage{natbib}
\usepackage{ulem}
\usepackage{subfigure}
\usepackage{graphicx}
\usepackage{CJKutf8}                   % see the JCAP-author-manual

\usepackage[T1]{fontenc} % if needed

\title{\boldmath lmage of a time-dependent rotating regular black hole}

\author[a]{Sen Guo,\note{Corresponding author.}}
\author[b]{En-Wei Liang}
\author[b]{Yu-Xiang Huang}
\author[c]{Yu Liang}
\author[d]{Qing-Quan Jiang}
\author[e]{Kai Lin}
\author[f]{Li-Fang Li,\note{Corresponding author.}}

\affiliation[a]{College of Physics and Electronic Engineering, Chongqing Normal University,\\Chongqing 401331, People's Republic of China}
\affiliation[b]{Guangxi Key Laboratory for Relativistic Astrophysics, School of Physical Science and Technology, Guangxi University,\\Nanning 530004, People's Republic of China}
\affiliation[c]{Hubei Subsurface Multi-scale Imaging Key Laboratory, School of Geophysics and Geomatics, China University of Geosciences,\\Wuhan 430074, People's Republic of China}
\affiliation[d]{School of Physics and Astronomy, China West Normal University,\\Nanchong 637000, People's Republic of China}
\affiliation[e]{Universidade Federal de Campina Grande, Campina Grande,\\PB 58429-900, Brasil}
\affiliation[f]{Center for Gravitational Wave Experiment, National Microgravity Laboratory, Institute of Mechanics, Chinese Academy of Sciences,\\Beijing 100190, People's Republic of China}

\emailAdd{sguophys@126.com}
\emailAdd{lilifang@imech.ac.cn}

\abstract{In this study, we develop a modeling framework based on spatio-temporal generalized random fields to simulate the time-evolving accretion flows and their associated imaging signatures around rotating regular \textcolor{black}{black holes}. We extend the Matérn field formalism to the spatio-temporal domain and introduce a locally anisotropic tensor structure \(\Lambda(\mathbf{x})\), which encodes direction-dependent correlation scales motivated by Keplerian velocity fields, thereby generating physically informed perturbation structures. Coupled with a \textcolor{black}{computationally efficient} light ray-tracing scheme, this framework produces a sequence of time-resolved images of regular \textcolor{black}{black hole shadow and accretion structures}. By incorporating light-travel time effects, we identify significant temporal smearing of \textcolor{black}{features within} strongly lensed regions and rapidly varying sources, thus enhancing the physical realism of the modeling. Comparison with existing general relativistic magnetohydrodynamic simulations demonstrates that our stochastic generative model maintains statistical consistency while offering substantial computational efficiency. Moreover, the simulated results reproduce the dynamic positional shift of the bright ring structure observed in M87$^{*}$, providing theoretical support for interpreting its time-variable images.}

\begin{document}
\maketitle
\flushbottom

\section{Introduction}
\label{sec:intro}
Black holes (BHs), as extreme manifestations of spacetime curvature, give rise to a distinct observational signature known as the \textit{shadow}—a central dark region surrounded by lensed emission from infalling matter~\cite{1,2}. The theoretical foundations of BH shadow formation were first established by Luminet~\cite{3}, who employed ray-tracing techniques to model the optical appearance of Schwarzschild BHs, revealing asymmetric brightness patterns driven by strong-field gravitational lensing. Subsequent investigations in the context of Kerr spacetimes demonstrated that the shape and size of the shadow are governed primarily by the BH's spin and observer inclination, providing a means to probe the underlying geometry \textcolor{black}{with minimal dependence on} the astrophysical environment~\cite{4,5}. Analytical studies further uncovered that the photon ring—a narrow, self-similar substructure near the shadow edge—encodes \textcolor{black}{fundamental and potentially universal} information about the spacetime~\cite{6}. These theoretical predictions have since been substantiated by very-long-baseline interferometry (VLBI) observations conducted by the Event Horizon Telescope (EHT), which successfully imaged the shadows of M87$^*$~\cite{7,8} and Sagittarius A$^*$ (Sgr A$^*$)~\cite{9}. Nevertheless, such observations also underscore the necessity to incorporate \textcolor{black}{accretion flow variability and radiative transfer effects}, and to explore non-Kerr spacetimes, including \textcolor{black}{regular and horizonless BH alternatives}, to fully interpret high-resolution image structures~\cite{10,11}.

\par
The advent of the EHT observations of M87$^*$ and Sgr A$^*$ has ushered in a new era of precision BH imaging, necessitating theoretical models that self-consistently incorporate both the spacetime geometry and the complexities of turbulent accretion flows. General relativistic magnetohydrodynamic (GRMHD) simulations have demonstrated that while short-term variability in the accretion process tends to obscure photon ring structures in individual snapshots, their signatures persist in time-averaged images, enabling stringent tests of \textcolor{black}{Einstein's theory of general relativity} in the strong-field regime~\cite{8,12}. However, the majority of these studies adopt the Kerr metric as the underlying spacetime, thereby neglecting potential deviations from classical \textcolor{black}{predictions of general relativity}. Such deviations may arise in a variety of theoretical contexts, including quantum gravity-inspired regular BH models featuring nonsingular cores~\cite{13,14}, modifications of the gravitational action~\cite{15}, or emergent spacetime geometries derived from \textcolor{black}{higher-dimensional or string-theoretic frameworks}~\cite{16}. There are also some excellent discussions about black hole images that appear on ~\cite{aa1,aa2,aa3,aa4,aa5,aa6,aa7,aa8,aa9,aa10}. Recent analyses have shown that these non-Kerr spacetimes can produce distinctive observational imprints—such as suppressed inner shadow regions, centroid shifts, and azimuthal brightness asymmetries—within VLBI-resolved images~\cite{11,17,18}. Nevertheless, such signatures remain largely unexplored in fully time-dependent GRMHD models, underscoring the need for further investigation into the interplay between accretion dynamics and non-Kerr geometries.

\par
A significant theoretical gap remains in the modeling of shadows cast by regular BHs—nonsingular alternatives to the classical Kerr solution. The Hayward metric~\cite{13}, characterized by a finite-curvature core, has been shown to produce slightly smaller shadow diameters compared to Kerr BHs, primarily due to \textcolor{black}{repulsive quantum-gravity-induced effects} near the center~\cite{14}. However, early studies were confined to static, non-rotating configurations, neglecting the essential roles of spin and dynamical accretion processes, thereby limiting their direct observational applicability. Recent efforts have extended these investigations to rotating Hayward spacetimes~\cite{19}, identifying \textcolor{black}{non-negligible} displacements in photon ring structures. Nevertheless, such analyses often disregard key physical effects, including the influence of magnetic charge and \textcolor{black}{temporal asymmetries in emission}. Notably, existing studies have yet to resolve the degeneracy between spin and magnetic charge—both of which contribute comparably to shadow size modifications in static metrics~\cite{17}—nor have they incorporated stochastic variability intrinsic to turbulent accretion flows~\cite{7,8,9,10}. For example, Tsukamoto~\cite{20} demonstrated that magnetic charge alters photon capture radii in regular BHs, yet this influence remains unexplored in \textcolor{black}{fully time-dependent} ray-tracing scenarios. Similarly, while Cunha et al.~\cite{21} developed geometric diagnostics to disentangle spin and charge effects in Kerr–Newman BHs, analogous tools have not been established for regular spacetimes. Recent GRMHD simulations by Younsi et al.~\cite{22} further underscore the impact of turbulent accretion on temporal shadow morphology, revealing variability signatures inaccessible to static models. These findings collectively emphasize the necessity of \textcolor{black}{self-consistent, time-resolved} analyses within the framework of non-Kerr and regular BH geometries.

\par
The temporal evolution of BH shadows presents a significant challenge for both theoretical modeling and observational interpretation. While time-averaged GRMHD simulations robustly confirm the persistence of photon rings as stable relativistic signatures~\cite{8}, short-timescale variability in the accretion flow induces transient fluctuations in brightness and morphology~\cite{23,24}. This variability complicates image reconstruction efforts by future space-based VLBI arrays, which are constrained by sparse and irregular sampling in both time and Fourier domains~\cite{25,26}. Accurate interpretation therefore requires predictive frameworks capable of capturing the \textcolor{black}{full spatiotemporal} evolution of BH shadow structures. For example, Chael et al.~\cite{27} demonstrated that stochastic turbulence in the accretion disk imprints frequency-dependent signatures onto interferometric visibilities, while Blackburn et al.~\cite{28} quantified the systematic biases introduced by incomplete temporal sampling on inferred shadow geometries. Emerging time-resolved ray-tracing techniques---such as those incorporating \textcolor{black}{spatio-temporal Gaussian Markov random fields (GMRFs)}~\cite{29}---show promise in modeling such effects, although their extension to non-Kerr spacetimes remains an \textcolor{black}{important and largely uncharted} frontier.

\par
This work will investigate \textcolor{black}{the first fully time-resolved analysis} of dynamical shadows in magnetically charged, rotating Hayward BHs. Building upon our previous derivation of null geodesics in rotating BH spacetimes~\cite{18}, we implement a ray-tracing scheme coupled with a novel framework, \textsc{INOISY}, to model time-dependent radiation from inhomogeneous and anisotropic Gaussian random fields (GRFs). This methodology allows for a detailed analysis of the temporal evolution of BH images, capturing transient variations in shadow and photon ring morphology induced by stochastic emission processes. Our findings offer \textcolor{black}{quantitative} insights into the coupling between relativistic light propagation and turbulent accretion dynamics in non-Kerr geometries. These developments provide a \textcolor{black}{robust} theoretical foundation for testing quantum gravitational signatures through the evolving morphology of BH shadows, thereby bridging key theoretical and observational challenges in strong-field gravity.

\par
The structure of this work is as follows: In Sec.\ref{sec:2}, we review the spacetime geometry of rotating Hayward BHs and develop an efficient stochastic framework based on anisotropic and inhomogeneous GRFs to model time-dependent accretion-driven emission structures in their vicinity. Sec.~\ref{sec:3} introduces a spatio-temporal imaging framework that integrates anisotropic Matérn fields with both fast-light ray tracing to model and visualize dynamic brightness fluctuations in accretion disks around rotating Hayward BHs. Finally, Sec.~\ref{sec:4} provides conclusions and a discussion of future directions.

\section{Rotating Hayward BH and Gaussian random fields}
\label{sec:2}
\par
In this section, we \textcolor{black}{provide a concise overview} of the rotating Hayward BH, \textcolor{black}{focusing on the spacetime geometry and geodesic structure} previously examined in our earlier work. The metric describing the rotating Hayward BH is given by~\cite{29}:
\begin{eqnarray}
\label{1}
{\rm d}s^2 = && - \Big(1 - \frac{2 m(r) r}{\Sigma} \Big) {\rm d} t^2 - \frac{4 a m(r) r}{\Sigma} \sin^{2}\theta {\rm d} t {\rm d} \phi + \frac{\Sigma}{\Delta} {\rm d} r^2  \nonumber\\
&&  + \Sigma {\rm d} \theta^{2} + \Big(r^2 + a^2 +\frac{2 m(r) r a^2 \sin^{2}\theta}{\Sigma} \Big) \sin^{2}\theta {\rm d} \phi^{2},
\end{eqnarray}
where
\begin{equation}
\label{2}
\Sigma = r^{2} + a^{2}\cos^{2}\theta,~~~~~~\Delta = r^{2} + a^{2} - 2 m(r) r.
\end{equation}
Here, $m(r)$ represents the mass function, defined as~\cite{13}
\begin{equation}
\label{3}
m(r) = \frac{M r^3}{r^3 + g^3},
\end{equation}
where $a$ denotes the BH spin and $g$ corresponds to the magnetic charge. In the limiting case $g \rightarrow 0$, the rotating Hayward BH reduces to the Kerr BH, \textcolor{black}{recovering the causal structure characteristic of} the Kerr spacetime within the framework of the rotating Hayward BH.

\par
The real astrophysical BHs are immersed in highly dynamic environments, characterized by time-variable accretion disks and evolving plasma flows. To address this discrepancy, we extend our framework to incorporate time-dependent accretion dynamics within the rotating Hayward BH spacetime. We investigate how transient astrophysical processes modulate the morphology of BH shadows and photon rings. This extension enables a more realistic assessment of the observational signatures arising from non-stationary emission near regular rotating BHs. In both astronomical observations and theoretical modeling, variable sources exhibiting non-stationary and non-axisymmetric characteristics are generally categorized into two distinct classes. The first class comprises deterministic phenomena governed by well-defined physical equations, such as orbiting ``hot spots''—localized regions of enhanced emissivity that circle the central BH and produce characteristic light echo patterns~\cite{31,32,33,34}. However, the transient and short-lived nature of hot spots poses substantial observational challenges, as highlighted in the case of M87$^*$. The second class encompasses stochastic variability, wherein turbulent plasma in the vicinity of the BH may generate photon ring–like features or structures resembling hot spots. These signatures are often attributed to \textcolor{black}{inherent statistical fluctuations in the emissivity field}. Consequently, such stochastic variability can be exploited to \textcolor{black}{extract ensemble-averaged constraints on} the underlying morphology and dynamics of the emitting region.

\par
In this context, the accretion disk surrounding the BH may exhibit surface brightness fluctuations analogous to those observed in the disk around M87$^*$. Although numerical solutions to the governing magnetohydrodynamic equations are often employed to quantify the amplitude and structure of these fluctuations, fully three-dimensional simulations remain computationally prohibitive for many applications. To circumvent this limitation, GRFs have been widely adopted in astrophysical modeling as an efficient surrogate for simulating stochastic emission and noise patterns~\cite{35}. We introduce a modeling framework that incorporates surface brightness fluctuations within the accretion disk, represented by a GRF, to investigate the dynamical properties of rotating Hayward BHs. Accurately characterizing the spatial and temporal variability of the disk requires accounting for fluctuations that depend sensitively on the local radius—features that are inherently difficult to capture using standard Fourier-based techniques due to the disk’s inhomogeneous structure~\cite{36}. Moreover, the surface brightness exhibits anisotropic spatial correlations, further complicating conventional modeling approaches. To address these challenges, we construct an \textcolor{black}{inhomogeneous, anisotropic GRF} tailored to simulate realistic brightness fluctuations in the accretion flows surrounding rotating Hayward BHs. This modeling approach is supported by the central limit theorem, which states that the superposition of a large number of independent and identically distributed random processes tends toward a Gaussian distribution. This principle underpins the widespread use of GRFs in cosmology, where primordial perturbations are often treated as \textcolor{black}{statistically homogeneous and isotropic scalar fields} generated during the inflationary epoch.

\par
To construct an \textcolor{black}{inhomogeneous, anisotropic Matérn random field}, we \textcolor{black}{employ geostatistical methods by solving the stochastic partial differential equation (SPDE)}~\cite{37}. \textcolor{black}{This approach} yields a spatial field with a \textcolor{black}{rigorously defined local} covariance structure. Specifically, the field statistics are governed by the SPDE whose Green’s function corresponds to the Matérn covariance, characterized by a power spectrum that remains flat on large spatial scales and decays as a power law at smaller scales. \textcolor{black}{Such behavior} captures both long-range correlations and localized variability, making it ideally suited for modeling structured turbulence in astrophysical flows. \textcolor{black}{For homogeneous, isotropic GRFs, sampling can be efficiently performed via Fourier transform methods.} In this context, isotropy implies that the covariance function depends solely on the spatial separation \(\Delta x\). The power spectrum is defined as~\cite{36}:
\begin{eqnarray}
P(\mathbf{k}) &=& \langle \hat{h}(k)\,\hat{h}^*(k') \rangle \nonumber \\
&=& \int \langle h(x)\,h^*(x') \rangle\,e^{i(kx - k'x')} \,\mathrm{d}x\,\mathrm{d}x' \nonumber \\
&=& \int C(\Delta x)\,e^{i(k - k')x}\,e^{ik'\,\Delta x} \,\mathrm{d}x\,\mathrm{d}\Delta x
\quad \Delta x \equiv x - x' \nonumber \\
&=& 2\pi\,\hat{C}(k')\,\delta(k - k'),
\label{4}
\end{eqnarray}
where $\langle h(x) \rangle$ denotes the ensemble average, and the Fourier transform of the spatial field $h(x)$ is given by $\hat{h}(k) = \int h(x)\, e^{ikx}\, \mathrm{d}x$. Accordingly, the power spectrum $P(\mathbf{k})$ is the Fourier transform of the covariance function $C(\Delta x)$.

\par
To \textcolor{black}{enable numerical implementation of these stochastic fields}, we \textcolor{black}{adopt a discretized GMRF formulation} on a regular lattice. Unlike continuous GRFs, GMRFs are specified not by their covariance matrix \(\mathbf{C}\) but by the \textit{precision matrix} \(\mathbf{Q} = \mathbf{C}^{-1}\). The precision matrix is typically sparse due to the \textit{Markov property}, which enforces conditional independence between non-neighboring grid points. In particular, the joint probability density for a zero-mean GMRF is given by:
\begin{equation}
\label{5}
p(\mathbf{h}) = \frac{1}{(2\pi)^{n/2}\,\lvert \mathbf{Q}\rvert^{-1/2}} \exp\!\biggl(-\tfrac{1}{2}\,\mathbf{h}^\top \mathbf{Q} \,\mathbf{h}\biggr),
\end{equation}
where \(\mathbf{h}\in\mathbb{R}^n\) denotes the discretized field vector and \(n\) is the total number of grid points. The \textcolor{black}{sparse, banded structure} of \(\mathbf{Q}\) enables efficient Cholesky factorization and sampling, \textcolor{black}{substantially reducing computational cost} for high-resolution astrophysical simulations.

\par
In \textcolor{black}{astrophysical modeling}, GMRFs can be \textcolor{black}{constructed} by numerically solving SPDEs, \textcolor{black}{providing} fine-grained control over local covariance structures. A \textcolor{black}{widely adopted} formulation links a Matérn-type random field to the linear SPDE~\cite{38}:
\begin{equation}
\label{6}
(\kappa^2 - \Delta)^{\alpha/2} h(x) = \mathcal{W}(x),
\end{equation}
where \(\kappa > 0\) \textcolor{black}{sets} the correlation length, \(\Delta\) is the Laplacian operator, \(\alpha\) \textcolor{black}{controls} the field smoothness, and \(\mathcal{W}(x)\) denotes Gaussian white noise. Discretization of this SPDE yields a sparse precision matrix \(\mathbf{Q}\) via \textit{\textcolor{black}{finite-difference methods}} on a structured grid or \textit{\textcolor{black}{finite-element methods}} on an unstructured mesh. In the former, \textcolor{black}{central difference} approximations produce a banded linear system whose inverse recovers the Matérn covariance, and \(\mathbf{Q}\) inherits the local stencil sparsity. In the latter, \textcolor{black}{variational} discretization over triangular elements affords geometric flexibility, \textcolor{black}{advantageous} for curved or multi-scale astrophysical domains. For finite-element schemes, one obtains
\begin{equation}
\label{7}
\mathbf{Q} = \kappa^4 \mathbf{C} + 2\kappa^2 \mathbf{G} + \mathbf{G} \mathbf{C}^{-1} \mathbf{G},
\end{equation}
where \(\mathbf{C}\) and \(\mathbf{G}\) are the mass and stiffness matrices, respectively. These \textcolor{black}{sparse, symmetric positive-definite} precision matrices enable efficient Cholesky factorization and \textcolor{black}{scalable} sampling for high-resolution astrophysical simulations.

\par
We consider the \textcolor{black}{stationary, isotropic Matérn GRF} \(\mathcal{F}_{\nu}(\mathbf{x})\) of smoothness order \(\nu\) and zero mean. This field follows the Matérn covariance structure, defined by
\begin{equation}
\label{8}
\mathbf{C}(\mathbf{x}, \mathbf{y}) = \frac{\sigma^{2}}{2^{\nu - 1}\,\Gamma(\nu)} 
\left(\frac{|\mathbf{x} - \mathbf{y}|}{\lambda}\right)^{\!\nu}
K_{\nu}\!\Bigl(\frac{|\mathbf{x} - \mathbf{y}|}{\lambda}\Bigr),
\end{equation}
where \(\Gamma(\cdot)\) denotes the Gamma function, \(\nu\) is the smoothness parameter controlling the \textcolor{black}{differentiability} of the field, \(\lambda\) is the \textcolor{black}{correlation scale}, and \(K_{\nu}\) is the modified Bessel function of the second kind. For the special case \(\nu = 2 - \tfrac{d}{2}\), the Matérn field can be realized as the solution to the linear SPDE~\cite{38}:
\begin{equation}
\label{9}
(1 - \lambda^{2}\,\nabla^{2})\,\mathcal{F}_{\nu}(\mathbf{x}) 
= \mathcal{N}\,\sigma(\mathbf{x})\,\lambda^{d/2}\,\mathcal{W}(\mathbf{x}),
\end{equation}
where \(\mathcal{N}\) is a \textcolor{black}{normalization constant}, \(\sigma^{2}\) is the field variance, and \(\mathcal{W}(\mathbf{x})\) represents Gaussian white noise. The corresponding power spectrum for this formulation is given by
\begin{equation}
\label{10}
P(k) = \frac{\mathcal{N}^{2}\,\sigma^{2}\,\lambda^{d}}{\bigl[1 + (k\,\lambda)^{2}\bigr]^{2}},
\end{equation}
which exhibits a \textcolor{black}{flat plateau} at low wavenumbers and a \textcolor{black}{power-law decay} \(\propto k^{-4}\) at high wavenumbers, consistent with the Matérn class.

\par
The SPDE formulation for GRFs can be systematically extended to accommodate \textcolor{black}{spatially inhomogeneous} and \textcolor{black}{locally} anisotropic structures. This \textcolor{black}{generalized} framework enables the construction of \textcolor{black}{nonstationary} Matérn fields by introducing spatial variability into the model parameters~\cite{37}. To define an inhomogeneous and anisotropic Matérn field in \( d \)-dimensional space, we introduce a family of \( d \) orthonormal direction vectors \(\mathbf{u}_{l}(\mathbf{x})\) and their associated correlation lengths \(\lambda_{l}(\mathbf{x})>0\), for \(l=1,\ldots,d\). We then define the anisotropy matrix as:
\begin{equation}
\label{11}
\Lambda(\mathbf{x}) = \sum_{l=1}^{d} \lambda_{l}^{2}(\mathbf{x})\,\mathbf{u}_{l}(\mathbf{x})\,\mathbf{u}_{l}^{\top}(\mathbf{x}), 
\quad |\Lambda(\mathbf{x})| \equiv \det\Lambda(\mathbf{x}) = \prod_{l=1}^{d}\lambda_{l}^{2}(\mathbf{x}).
\end{equation}
The positive-definite matrix \(\Lambda(\mathbf{x})\) defines a Riemannian metric over \(\mathbb{R}^d\), with the corresponding line element
\begin{equation}
\label{12}
\mathrm{d}s^{2}(\Delta \mathbf{x};\mathbf{x}) 
= \Delta \mathbf{x}^{\top}\Lambda^{-1}(\mathbf{x})\,\Delta \mathbf{x}
= \sum_{l=1}^{d} \biggl(\frac{\Delta \mathbf{x}\cdot\mathbf{u}_{l}(\mathbf{x})}{\lambda_{l}(\mathbf{x})}\biggr)^{2}.
\end{equation}
This metric generalizes the notion of spatial separation in the Matérn covariance, yielding
\begin{equation}
\label{13}
C_{\nu}(\mathbf{x},\mathbf{x}') 
= \frac{1}{2^{\nu-1}\Gamma(\nu)}\,\mathrm{d}s^{\nu}(\mathbf{x},\mathbf{x}')
\,K_{\nu}\!\bigl(\mathrm{d}s(\mathbf{x},\mathbf{x}')\bigr),
\end{equation}
where \(K_{\nu}\) is the modified Bessel function of the second kind. The resulting anisotropic, nonstationary Matérn field satisfies the generalized SPDE:
\begin{equation}
\label{14}
\bigl(1-\nabla\cdot[\Lambda(\mathbf{x})\nabla]\bigr)\mathcal{F}(\mathbf{x}) 
= \mathcal{N}\,\sigma(\mathbf{x})\,|\Lambda(\mathbf{x})|^{1/4}\,\mathcal{W}(\mathbf{x}),
\end{equation}
with \(\mathcal{N}\) a normalization constant, \(\sigma^{2}(\mathbf{x})\) the local variance function, and \(\mathcal{W}(\mathbf{x})\) spatial white noise. This \textcolor{black}{formulation} provides a \textcolor{black}{versatile and computationally tractable} framework for simulating structured stochastic fields in anisotropic astrophysical environments.

\section{Time-Dependent Imaging and Dynamic Evolution}
\label{sec:3}
\par
As discussed in preceding section, we \textcolor{black}{modeled} the spatial structure of brightness fluctuations within the Matérn framework, \textcolor{black}{integrating} covariance kernels, SPDEs, and GMRFs. This \textcolor{black}{formalism} enables physically motivated and statistically \textcolor{black}{rigorous} simulation of both stationary and nonstationary fields. In practice, time‑averaged GRMHD images of astrophysical sources are often approximated by fixed, axisymmetric, equatorial emission profiles obtained via ray‑tracing methods~\cite{39}, which \textcolor{black}{provide} a computationally efficient strategy for modeling long‑term averaged or quasi‑static emission. To generate such profiles, we define the source intensity function as~\cite{40}:
\begin{equation}
\label{15}
I_{\mathrm{s}}\bigl(r_{\mathrm{s}}^{(n)}, \phi_{\mathrm{s}}^{(n)}, t_{\mathrm{s}}^{(n)}\bigr)
= J\bigl(r_{\mathrm{s}}\bigr),
\end{equation}
where the radial intensity profile \(J(r_{\mathrm{s}})\) is \textcolor{black}{specified analytically}~\cite{6} by
\begin{equation}
\label{16}
J(r; \mu, \vartheta, \gamma) \equiv 
\frac{\exp\!\Bigl[-\tfrac12\bigl(\gamma + \arcsin\tfrac{\gamma - \mu}{\vartheta}\bigr)^2\Bigr]}
{\sqrt{(r - \mu)^2 + \vartheta^2}}.
\end{equation}

\par
This intensity profile is derived from Johnson’s standard unbounded (SU) distribution and \textcolor{black}{admits} efficient numerical evaluation. The parameters \(\mu\), \(\vartheta\), and \(\gamma\) respectively control the \textcolor{black}{location}, \textcolor{black}{scale}, and \textcolor{black}{skewness} of the emission peak. Denoting the source coordinate by \(\mathbf{x}=(r_{\rm s},\phi_{\rm s},t_{\rm s})\), we introduce brightness fluctuations by augmenting Eq.~\eqref{16} with a perturbation term:
\begin{equation}
\label{17}
I_{\mathrm{s}}(\mathbf{x}_{\mathrm{s}})
=J(r_{\mathrm{s}})+\Delta J(\mathbf{x}_{\mathrm{s}})
=J(r_{\mathrm{s}})\bigl[1 + F(\mathbf{x}_{\mathrm{s}})\bigr],
\end{equation}
where \(F(\mathbf{x}_{\mathrm{s}})\) is a stochastic modulation field representing local deviations from axisymmetry. By construction, the ensemble mean satisfies
\begin{equation}
\label{18}
\langle I_{\mathrm{s}}(\mathbf{x}_{\mathrm{s}})\rangle
=\langle J(r_{\mathrm{s}})\rangle+\langle J(r_{\mathrm{s}})F(\mathbf{x}_{\mathrm{s}})\rangle
=J(r_{\mathrm{s}}),
\end{equation}
ensuring that the time‑averaged image remains consistent with the static profile. To imprint spatial correlations, we model \(F(\mathbf{x}_{\mathrm{s}})\) as a normalized, zero‑mean Matérn field:
\begin{equation}
\label{19}
\hat{\mathcal{F}}(\mathbf{x})
=\frac{\mathcal{F}(\mathbf{x})-\langle\mathcal{F}(\mathbf{x})\rangle}
{\sqrt{\langle\mathcal{F}(\mathbf{x})^2\rangle-\langle\mathcal{F}(\mathbf{x})\rangle^2}},
\end{equation}
and express the perturbed intensity as
\begin{equation}
\label{20}
I_{\mathrm{s}}(\mathbf{x}_{\mathrm{s}})
\approx J(r_{\mathrm{s}})\,\exp\bigl[\sigma\,\hat{\mathcal{F}}(\mathbf{x})\bigr],
\end{equation}
where \(\sigma\) is a tunable parameter controlling the fluctuation amplitude. To \textcolor{black}{preserve} the ensemble mean given by Eq.~\eqref{20}, we adopt the normalized exponential form:
\begin{equation}
\label{21}
I_{\mathrm{s}}(\mathbf{x}_{\mathrm{s}})
=J(r_{\mathrm{s}})\,\exp\!\bigl[\sigma\,\hat{\mathcal{F}}(\mathbf{x})-\tfrac12\sigma^2\bigr].
\end{equation}

\subsection{Time-dependent Emission Modeling}
\label{sec:3-1}
\par 
To extend our emission framework to time‑dependent sources, we generalize the spatial Matérn field \(\mathcal{F}(\mathbf{x})\) into a spatio‑temporal field \(\mathcal{F}(\mathbf{x},t)\), where \(\mathbf{x}=(r_{\rm s},\phi_{\rm s})\) denotes the equatorial coordinates and \(t\) denotes time. The corresponding spatio‑temporal covariance structure is given by:
\begin{equation}
\label{22}
C(\mathbf{x},t;\mathbf{x}',t')
=\sigma^2\,\rho_s\bigl(\|\mathbf{x}-\mathbf{x}'\|\bigr)\,
\textcolor{black}{\rho_t\bigl(|t-t'|\bigr)},
\end{equation}
where \(\rho_s\) is a spatial Matérn kernel and \(\rho_t\) is a temporal correlation function, for example 
\(\rho_t(\tau)=\exp(-\tau/\tau_c)\), with \(\tau_c\) the characteristic correlation time. To synthesize realizations of \(\mathcal{F}(\mathbf{x},t)\), we solve the following spatio‑temporal SPDE:
\begin{equation}
\label{23}
\bigl(\partial_t + \kappa^2 - \nabla\!\cdot\![\Lambda(\mathbf{x})\nabla]\bigr)^{\!\beta}
\,\mathcal{F}(\mathbf{x},t)
=\mathcal{W}(\mathbf{x},t),
\end{equation}
where \(\mathcal{W}(\mathbf{x},t)\) is spatio‑temporal white noise, and \(\beta>0\) \textcolor{black}{controls} the temporal smoothness. This formulation \textcolor{black}{yields} temporally coherent yet spatially structured fluctuations in the accretion‑disk emission.

\par
To \textcolor{black}{facilitate} comparison between synthetic emission models and interferometric observations, we compute the complex visibility function \(V(\boldsymbol{u},t)\), defined as the two‑dimensional Fourier transform of the sky brightness distribution:
\begin{equation}
\label{24}
V(\boldsymbol{u},t)
=\int_{\mathbb{R}^2}I(\boldsymbol{x},t)\,\exp\bigl[-2\pi i\,\boldsymbol{u}\!\cdot\!\boldsymbol{x}\bigr]\,
\mathrm{d}^2\boldsymbol{x},
\end{equation}
where \(\boldsymbol{u}\) is the interferometric baseline vector and \(I(\boldsymbol{x},t)\) is the image‑plane intensity at coordinates \(\boldsymbol{x}\) and time \(t\). Assuming \textcolor{black}{an} axisymmetric brightness distribution such that \(I(\boldsymbol{x},t)=I(r,t)\) with \(r=|\boldsymbol{x}|\), Eq.~\eqref{24} reduces to a zeroth‑order Hankel transform:
\begin{equation}
\label{25}
V(u,t)=2\pi\int_{0}^{\infty}I(r,t)\,J_{0}(2\pi u r)\,r\,\mathrm{d}r,
\end{equation}
where \(u=|\boldsymbol{u}|\) and \(J_{0}\) is the Bessel function of the first kind. This \textcolor{black}{efficient} formulation enables direct comparison of time‑dependent synthetic models with observed complex visibilities from VLBI arrays such as the EHT and ngEHT.
\begin{figure}[htbp]
  \centering
  \includegraphics[width=13cm,height=9cm]{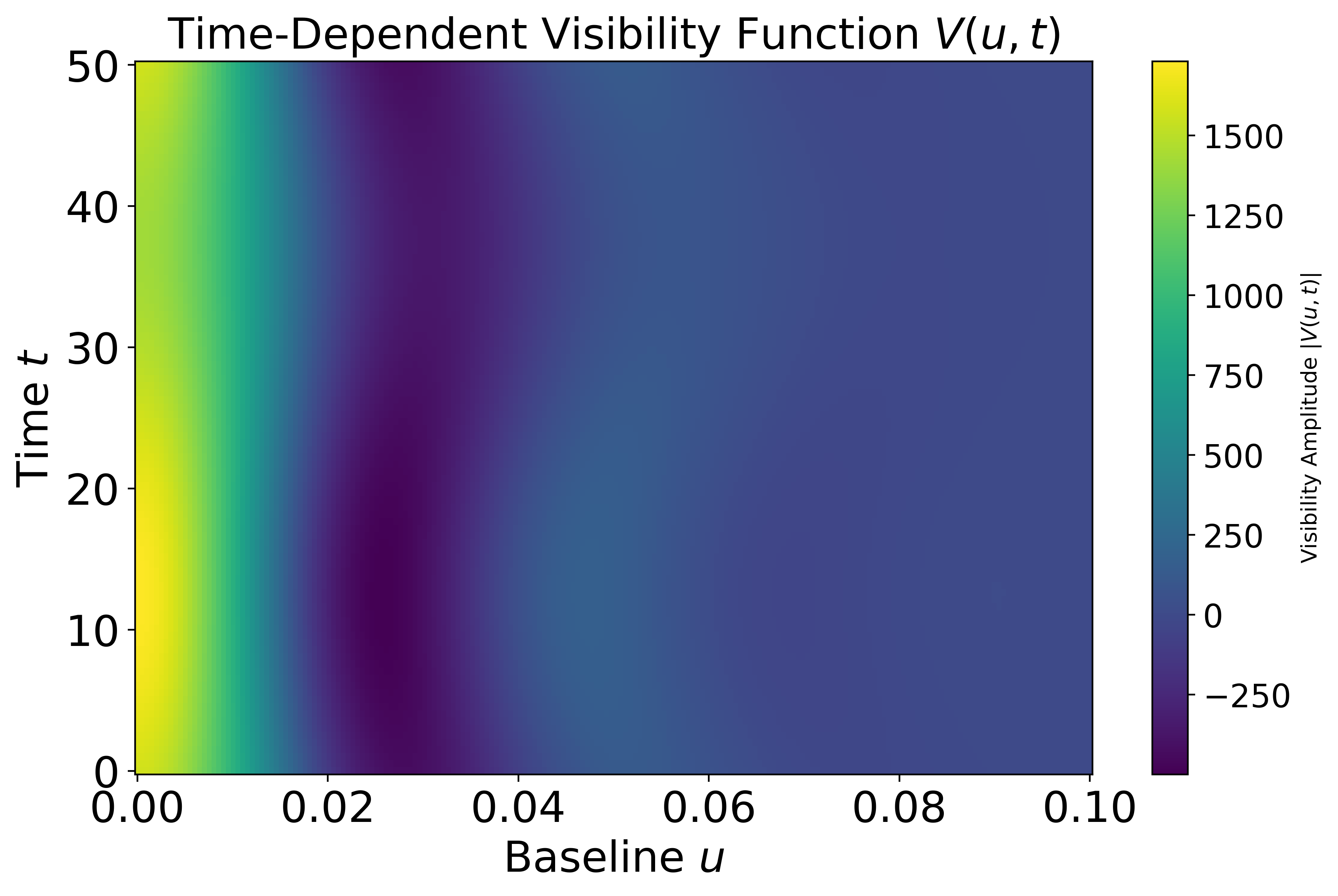}
  \caption{\textcolor{black}{Time‑Dependent Visibility Function \(V(u,t)\).}}
  \label{fig:1}
\end{figure}

In Figure~\ref{fig:1}, the \textcolor{black}{oscillatory features} in \(V(u,t)\) directly reflect the temporal evolution of the emission structure. As a ring or hotspot drifts radially due to sinusoidal modulation, the radial intensity profile \(I(r,t)\) acquires time dependence, leading to coherent variations in its Fourier counterpart. In the visibility domain, these effects appear as quasi‑periodic modulations along the time axis at fixed baseline lengths \(u\). Such signatures encode the characteristic spatial scales and frequencies of dynamical processes in the accretion flow. Interferometric arrays like the EHT or ngEHT can exploit this temporal behavior to constrain models of turbulence, orbiting features, or quasi‑periodic oscillations near the event horizon.

\subsection{Simulation of accretion disk and time correlation based on \textmd{INOISY}}
\label{sec:3-2}
\par
As \textcolor{black}{shown} in Eq.~(\ref{11}), our \textcolor{black}{modeling framework} requires specification of the unit vector fields \(\mathbf{u}_l(\mathbf{x}_{\mathrm{s}})\) and their associated correlation lengths \(\lambda_l\) for \(l=0,1,2\), each selecting a \textcolor{black}{distinct principal direction or eigenmode}. These quantities define the \textcolor{black}{position-dependent} anisotropy tensor
\begin{equation}
\Lambda(\mathbf{x}_{\mathrm{s}})
= \sum_{l=0}^{2} \lambda_l^{-2}\,\mathbf{u}_l(\mathbf{x}_{\mathrm{s}})
\,\mathbf{u}_l^\intercal(\mathbf{x}_{\mathrm{s}}),
\end{equation}
which \textcolor{black}{encodes} the directional \textcolor{black}{correlation scales} of the field, introducing local elongation or compression in the spatial structure of stochastic fluctuations.

\par
\begin{figure}[htbp]
\centering
\includegraphics[width=10cm,height=6cm]{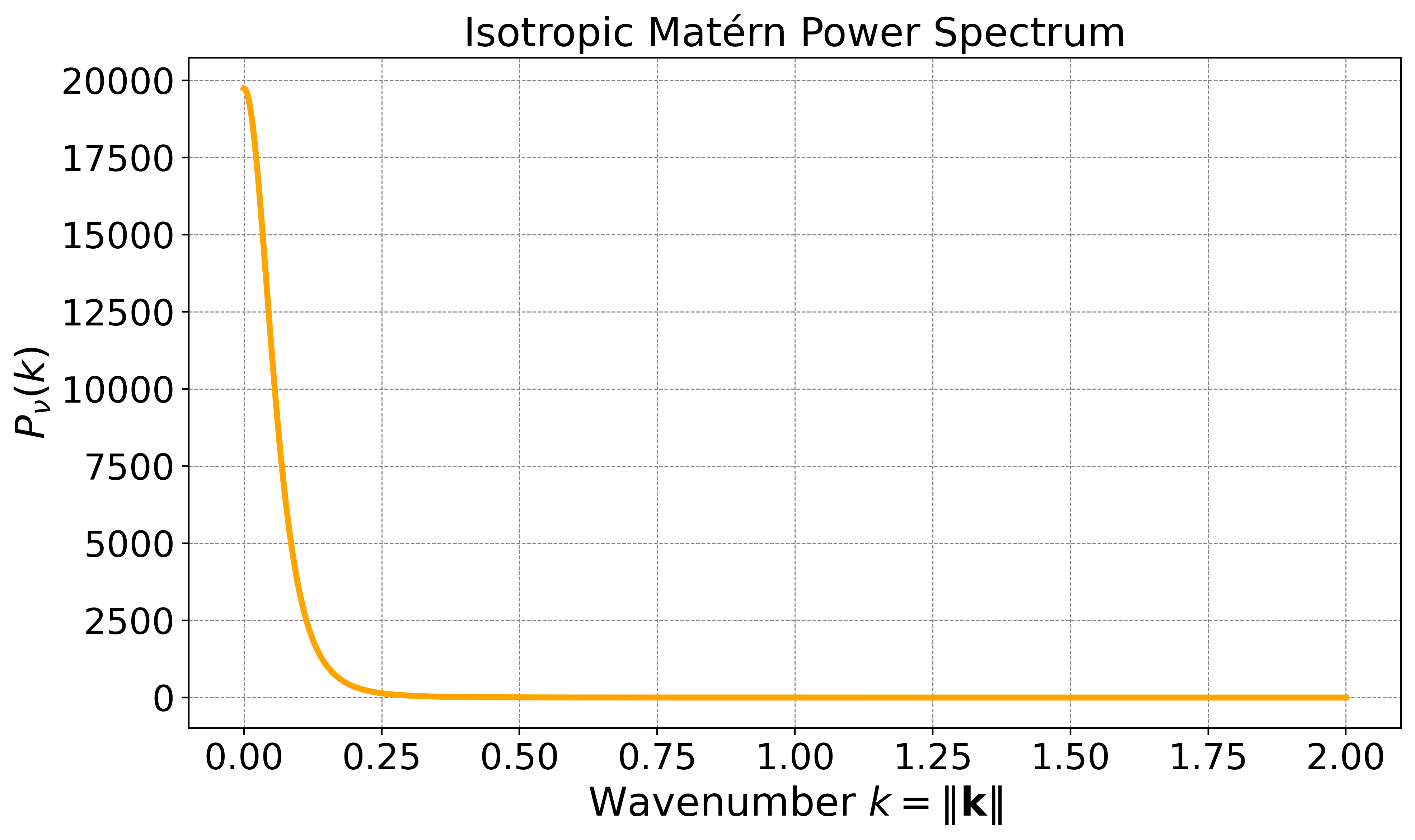}  
\caption{\textcolor{black}{Isotropic Matérn power spectral density \( P_\nu(k) \) as a function of wavenumber \( k \), shown for parameters \( \nu = 1.5 \), \( \kappa = 0.1 \), and \( \sigma^2 = 1.0 \). The spectrum decays as \( k^{-(2\nu + d)} \) at high frequencies, effectively controlling the small-scale roughness of the resulting field.}}
\label{fig:2}
\end{figure}

To \textcolor{black}{improve} the fidelity of synthetic images relative to realistic accretion flow structures, we model the log‑emissivity field \(\hat{\mathcal{F}}(\mathbf{x}_{\mathrm{s}})\) as a spatially inhomogeneous and anisotropic Matérn field. In the spectral domain, the power spectral density of the isotropic Matérn covariance is given by
\begin{equation}
P_{\nu}(\mathbf{k})
= \frac{\sigma^{2}\,\Gamma\!\bigl(\nu + \tfrac{d}{2}\bigr)}{\Gamma(\nu)}
\,\frac{(2\nu)^{\nu}}{(2\pi)^{d}\,\bigl(\kappa^{2}+\|\mathbf{k}\|^{2}\bigr)^{\nu + d/2}},
\end{equation}
where \(\nu>0\) is the smoothness parameter, \(d=2\) is the spatial dimension, \(\sigma^{2}\) the marginal variance, and \(\kappa\) the inverse correlation length, as shown in Figure~\ref{fig:2}. To incorporate anisotropy and spatial variability, we \textcolor{black}{replace} the Laplacian in the Matérn‑based SPDE with an elliptic operator featuring the spatially varying tensor \(\Lambda(\mathbf{x}_{\mathrm{s}})\). This \textcolor{black}{extension} yields a structured covariance aligned with the underlying flow geometry. The resulting field captures small‑scale turbulent features and directional coherence seen in GRMHD simulations~\cite{39}, thereby enabling the generation of \textcolor{black}{physically realistic} synthetic images for interferometric modeling.

\par
The Matérn power spectrum \textcolor{black}{displays} a characteristic decay rate governed by the smoothness parameter \(\nu\) and the inverse correlation length \(\kappa\). At low wavenumbers (large spatial scales), the spectrum remains approximately flat, indicating \textcolor{black}{strong large-scale coherence} in the emissivity distribution. In contrast, at high wavenumbers the power decays rapidly as a power law, \(\propto k^{-(2\nu + d)}\), \textcolor{black}{effectively suppressing small-scale, high-frequency fluctuations}. This \textcolor{black}{natural roll‑off} regulates the spatial roughness of the field, ensuring a \textcolor{black}{delicate balance} between global smoothness and localized structural features. By tuning \(\nu\) and \(\kappa\), one can emulate the \textcolor{black}{multiscale turbulent texture} characteristic of GRMHD simulations.
\begin{figure}[htbp]
\centering
\includegraphics[width=10cm,height=8cm]{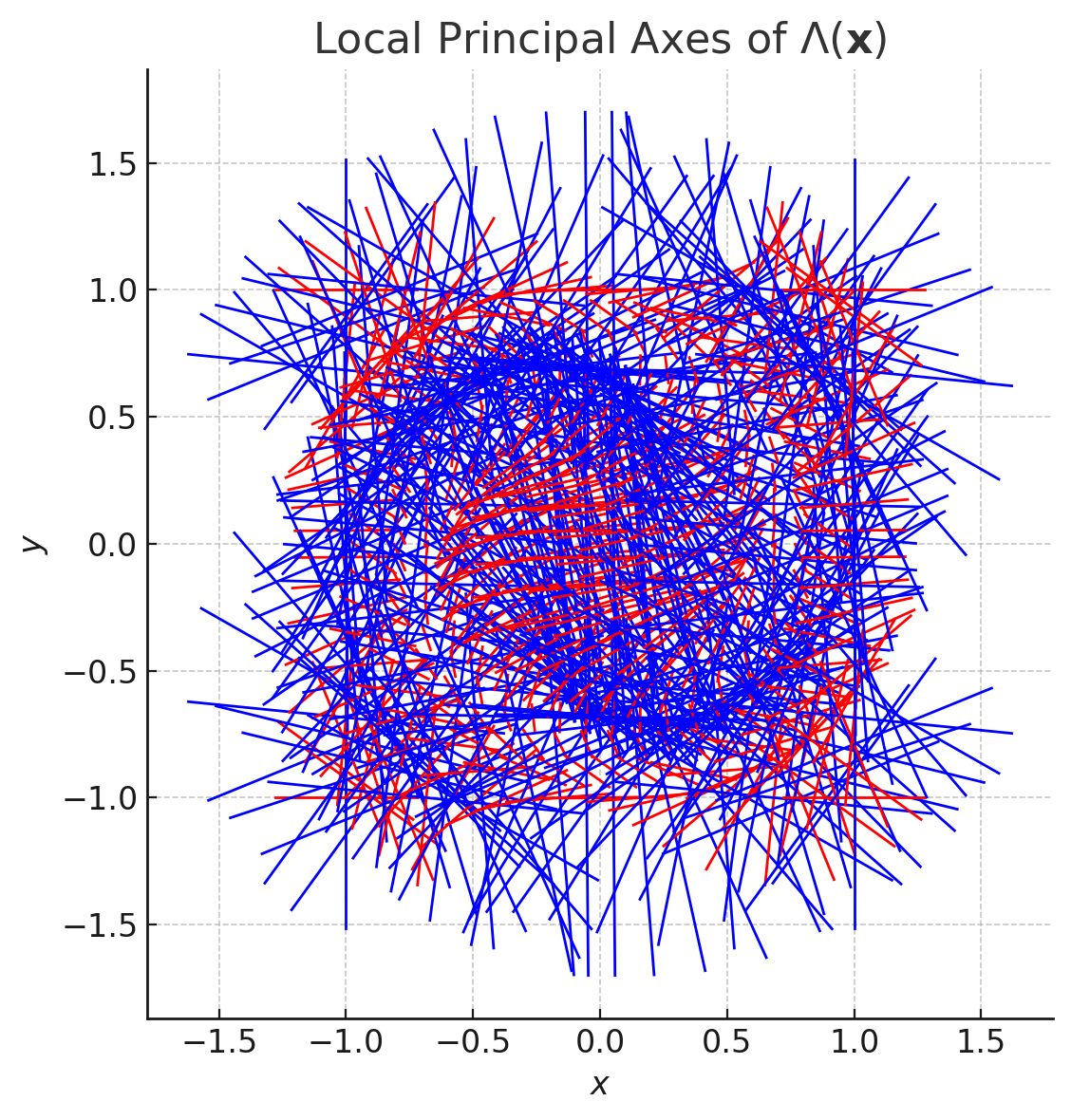}
\caption{\textcolor{black}{Local principal axes of the anisotropy tensor \( \Lambda(\mathbf{x}) \), visualized across a 2D spatial domain. Red segments denote the short axis directions, and black segments denote the long axis directions, each scaled by their corresponding inverse correlation lengths \( \lambda_l^{-1} \). The orientation and eccentricity vary smoothly with radius, mimicking the coherent but spatially inhomogeneous turbulence observed in GRMHD simulations.}}
\label{fig:3}
\end{figure}
\begin{figure}[htbp]
\centering
\includegraphics[width=10cm,height=8cm]{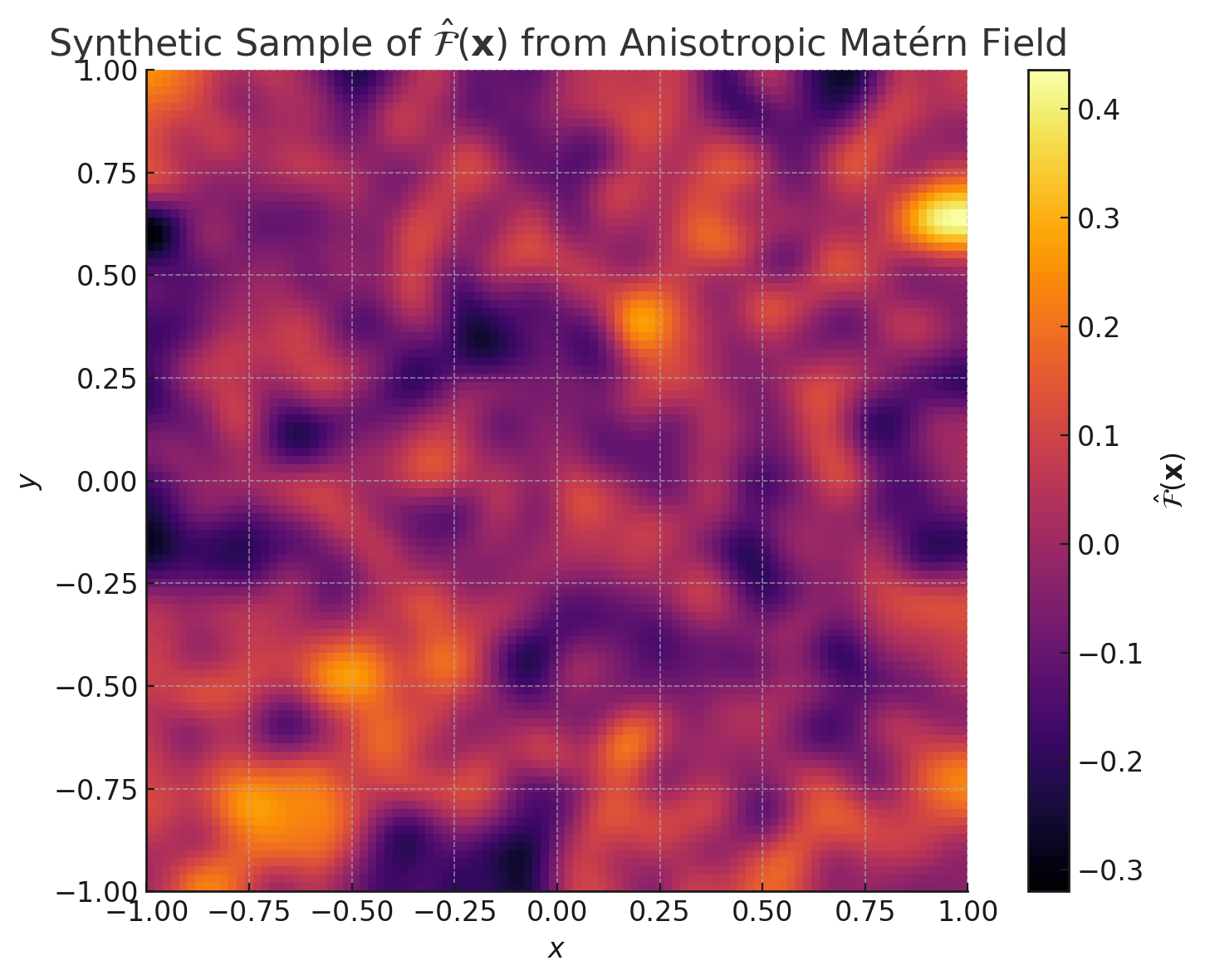}
\caption{\textcolor{black}{Synthetic realization of the log-emissivity field \( \hat{\mathcal{F}}(\mathbf{x}) \), generated from an inhomogeneous and anisotropic Matérn model with a spatially varying anisotropy tensor \( \Lambda(\mathbf{x}) \). The spatial structure exhibits smoothly modulated coherence scales and directionality, qualitatively reproducing the turbulent texture observed in GRMHD simulations.}}
\label{fig:4}
\end{figure}

\par
Figure~\ref{fig:3} \textcolor{black}{presents} the local principal axes of the anisotropy tensor \(\Lambda(\mathbf{x})\), with red and black line segments denoting the short‑axis and long‑axis directions, respectively. At each location in the 2D domain, these axes \textcolor{black}{reveal} the dominant orientation of spatial fluctuations and the associated correlation‑length anisotropy. Figure~\ref{fig:4} \textcolor{black}{displays} a synthetic realization of the normalized log‑emissivity field \(\hat{\mathcal{F}}(\mathbf{x})\) generated from the spatially varying tensor \(\Lambda(\mathbf{x})\). The resulting texture \textcolor{black}{exhibits} smoothly varying directionality and coherence scales. This simulation \textcolor{black}{approximates} the solution of a spatially inhomogeneous, anisotropic SPDE, thereby producing a nonstationary, anisotropic random field.

\par
We begin by analyzing the structure of the local covariance, which is governed by the symmetric positive‑definite anisotropy tensor \(\Lambda\), defined as
\begin{equation}
\label{28}
\Lambda = \lambda_{1}^{2}\,\mathbf{u}_{1}\,\mathbf{u}_{1}^{\mathsf{T}}
+ \lambda_{2}^{2}\,\mathbf{u}_{2}\,\mathbf{u}_{2}^{\mathsf{T}},
\end{equation}
where \(\lambda_{1}\) and \(\lambda_{2}\) denote the correlation lengths along the orthonormal directions specified by unit vectors \(\mathbf{u}_1\) and \(\mathbf{u}_2\), satisfying \(\mathbf{u}_{1}\cdot\mathbf{u}_{2}=0\). For a homogeneous field, the corresponding covariance function takes the form
\begin{equation}
\label{29}
C(\Delta \mathbf{x})
= \frac{\sigma^{2}}{2^{\nu-1}\,\Gamma(\nu)}\,
\mathrm{d}s^{\nu}(\Delta \mathbf{x})\,
K_{\nu}\!\bigl(\mathrm{d}s(\Delta \mathbf{x})\bigr),
\end{equation}
where \(\nu>0\) is the smoothness parameter, \(\Gamma(\cdot)\) is the gamma function, and \(K_{\nu}\) is the modified Bessel function of the second kind. The scaled Mahalanobis distance \(\mathrm{d}s(\Delta \mathbf{x})\) is given by
\begin{equation}
\label{30}
\mathrm{d}s^{2}(\Delta \mathbf{x})
= \Delta \mathbf{x}^{\mathsf{T}}\Lambda^{-1}\Delta \mathbf{x}
= \left(\frac{\Delta \mathbf{x}\cdot\mathbf{u}_{1}}{\lambda_{1}}\right)^{2}
+ \left(\frac{\Delta \mathbf{x}\cdot\mathbf{u}_{2}}{\lambda_{2}}\right)^{2}.
\end{equation}
\textcolor{black}{When \(\lambda_{1}=\lambda_{2}=\lambda\), this reduces to the standard isotropic Matérn covariance, recovering a homogeneous GRF. By contrast, setting \(\lambda_{1}\neq\lambda_{2}\) introduces directional dependence, and allowing \(\Lambda=\Lambda(\mathbf{x})\) endows the field with spatially varying anisotropy.}  
\begin{figure}[htbp]
\centering
\includegraphics[width=10cm,height=8cm]{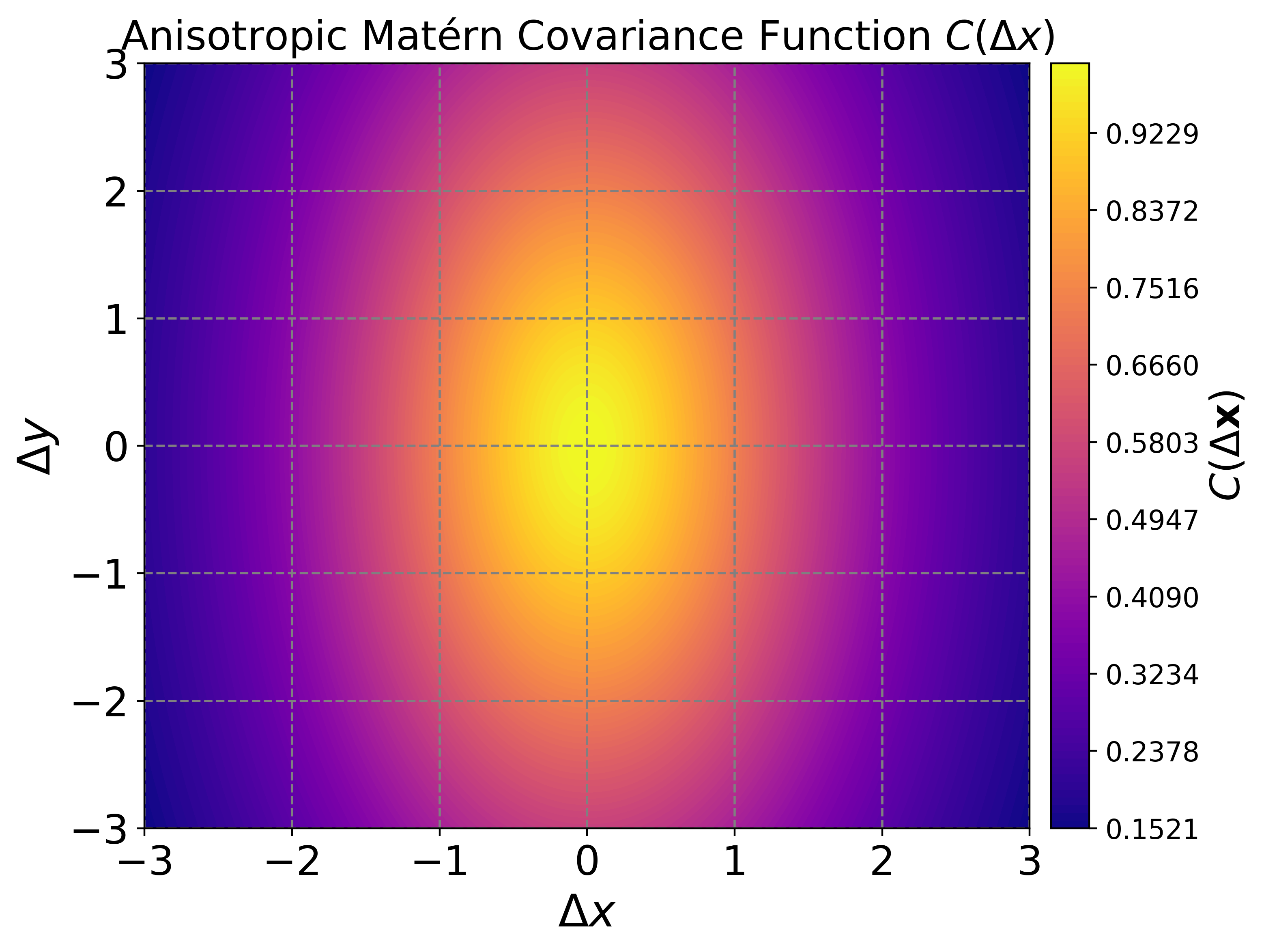}
\caption{\textcolor{black}{Spatial profile of the anisotropic Matérn covariance function \( C(\Delta \mathbf{x}) \) centered at the origin, computed for correlation lengths \( \lambda_1 = 1.0 \) and \( \lambda_2 = 2.0 \) along orthogonal directions. Elliptical contours reflect anisotropic decay of spatial correlation, with slower attenuation along the direction associated with larger correlation length.}}
\label{fig:5}
\end{figure}

\par
\begin{figure}[htbp]
\centering
\includegraphics[width=18cm,height=5cm]{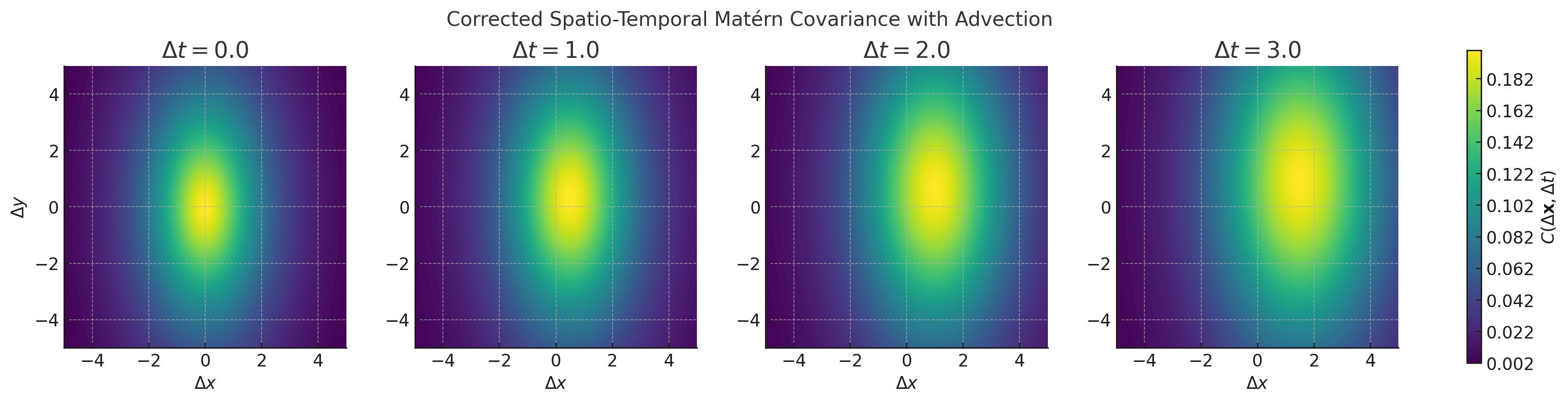}
\caption{\textcolor{black}{Time evolution of the spatio-temporal Matérn covariance function \( C(\Delta \mathbf{x}, \Delta t) \) under advection with velocity \( \mathbf{v} = (v_x, v_y) \). Each panel shows a spatial slice at fixed time lag \( \Delta t \). As time increases, the covariance peak shifts along the advection direction, illustrating coherent transport of emission features. The anisotropic spatial decay is preserved throughout the evolution.}}
\label{fig:6}
\end{figure}

To \textcolor{black}{incorporate} temporal variability, we extend our \textcolor{black}{spatially anisotropic} Matérn framework into a spatio‑temporal formulation by introducing two spatial dimensions and one temporal dimension. In this extended setting, the position vector is defined as \(\mathbf{x}=(t,x,y)\). Temporal evolution is \textcolor{black}{captured} via a constant velocity field \(\mathbf{v}=(0,v_x,v_y)\), representing advective transport in the spatial plane, together with a characteristic correlation time \(\lambda_0\). The generalized anisotropy tensor \(\Lambda(\mathbf{x}_{\rm s})\in\mathbb{R}^{3\times3}\) then comprises contributions from both temporal and spatial components:
\begin{equation}
\label{31}
\Lambda(\mathbf{x}_{\rm s})
= \lambda_0^2\,\mathbf{u}_0\mathbf{u}_0^{\mathsf{T}}
+ \lambda_1^2\,\mathbf{u}_1\mathbf{u}_1^{\mathsf{T}}
+ \lambda_2^2\,\mathbf{u}_2\mathbf{u}_2^{\mathsf{T}},
\end{equation}
where \(\mathbf{u}_0=(1,0,0)\) denotes the temporal axis, and \(\mathbf{u}_1,\mathbf{u}_2\) form an orthonormal basis in the spatial plane. Under this construction, the squared Mahalanobis distance between two spatio‑temporal points becomes
\begin{equation}
\label{32}
s^2
=\Bigl(\tfrac{\Delta t}{\lambda_0}\Bigr)^2
+\Bigl(\tfrac{(\Delta\mathbf{x}-\mathbf{v}\,\Delta t)\!\cdot\mathbf{u}_1}{\lambda_1}\Bigr)^2
+\Bigl(\tfrac{(\Delta\mathbf{x}-\mathbf{v}\,\Delta t)\!\cdot\mathbf{u}_2}{\lambda_2}\Bigr)^2,
\end{equation}
where \(\Delta\mathbf{x}=(x'-x,y'-y)\) and \(\Delta t=t'-t\). This formulation \textcolor{black}{naturally} introduces both temporal coherence and spatial advection into the effective correlation structure, enabling realistic modeling of evolving emission features in the image plane. Figure~\ref{fig:6} illustrates the spatio‑temporal Matérn covariance \(C(\Delta\mathbf{x},\Delta t)\) at several time lags \(\Delta t\). As \(\Delta t\) increases, the covariance peak shifts along the direction of \(\mathbf{v}=(v_x,v_y)\), \textcolor{black}{demonstrating} clear translational symmetry.

\par
As \textcolor{black}{specified} in Eq.~(\ref{31}), the unit vector \(\mathbf{u}_0\) \textcolor{black}{encodes} the temporal coherence of the accretion flow and is \textcolor{black}{parametrized} by the local correlation time \(\lambda_0(r_{\rm s})\). By contrast, the vectors \(\mathbf{u}_1\) and \(\mathbf{u}_2\) \textcolor{black}{delineate} the instantaneous spatial correlation orientations, associated with the correlation lengths \(\lambda_1(r_{\rm s})\) and \(\lambda_2(r_{\rm s})\), respectively. For spatio‑temporal displacements \(\Delta\mathbf{x}=(\Delta t,\Delta x,\Delta y)\), we introduce a triplet of basis vectors in \(\mathbb{R}^3\), comprising a flow‑aligned temporal axis and two orthonormal spatial axes:
\begin{align}
\label{33}
\mathbf{u}_0(\mathbf{x}_{\rm s}) 
&= \bigl(1,\,v_x(\mathbf{x}_{\rm s}),\,v_y(\mathbf{x}_{\rm s})\bigr), \\
\label{34}
\mathbf{u}_1(\mathbf{x}_{\rm s}) 
&= \bigl(0,\,\cos\theta(\mathbf{x}_{\rm s}),\,\sin\theta(\mathbf{x}_{\rm s})\bigr), \\
\label{35}
\mathbf{u}_2(\mathbf{x}_{\rm s}) 
&= \bigl(0,\,-\sin\theta(\mathbf{x}_{\rm s}),\,\cos\theta(\mathbf{x}_{\rm s})\bigr).
\end{align}
\textcolor{black}{Notably}, \(\mathbf{u}_1\) and \(\mathbf{u}_2\) remain orthonormal by construction (\(\mathbf{u}_1\cdot\mathbf{u}_2=0\)), \textcolor{black}{whereas} \(\mathbf{u}_0\) \textcolor{black}{need not} be orthogonal to the spatial basis. Since 
\(\lvert\Lambda(\mathbf{x}_{\rm s})\rvert = \lambda_0^2\lambda_1^2\lambda_2^2\), the tensor \(\Lambda(\mathbf{x}_{\rm s})\) remains positive‑definite, satisfying the requirements of Eq.~(\ref{11}). While \(\mathbf{u}_1\) and \(\mathbf{u}_2\) may be \textcolor{black}{arbitrarily oriented} to capture local spatial anisotropy, the temporal basis \(\mathbf{u}_0\) is \textcolor{black}{uniquely aligned} with the flow velocity, thereby embedding the temporal correlation structure directly into the model.

\par
By incorporating the time‑dependent GRF developed in the preceding sections into the \textmd{INOISY} framework, we construct a \textcolor{black}{physically motivated} synthetic accretion‑disk model that captures both spatial brightness fluctuations and temporal variability. \textcolor{black}{Guided} by observational and theoretical studies of the supermassive BH M87$^{*}$, we assign velocity vectors to fluid elements in the disk according to a Keplerian rotation profile~\cite{7,8,9,10}. This prescription \textcolor{black}{naturally} introduces advective transport and coherence timescales consistent with GRMHD simulations and VLBI constraints. We first construct the fractional fluctuation field \(f\), which encodes surface‑brightness perturbations governed by the spatio‑temporal covariance of Eq.~(\ref{31}). The velocity field is given by \( \mathbf{v} = \Omega_K\, \hat{\mathbf{z}} \times \mathbf{x} \) with \(\Omega_{K}\propto r^{-3/2}\). The principal axis \(\mathbf{u}_{1}\) of the spatial anisotropy tensor is \textcolor{black}{tilted} by \(20^\circ\) relative to the local circular orbit, thereby capturing the pitch angle of spiral patterns in the GRF, in agreement with local shearing‑box studies. To represent radial disk structure, we allow the correlation scales \(\lambda_{0}(r)\), \(\lambda_{1}(r)\), and \(\lambda_{2}(r)\) to vary with radius while remaining azimuthally uniform. Under the assumption \(H\propto r\) for the vertical scale height, we set \(\lambda_{1}\propto r\) and enforce a \textcolor{black}{fixed} anisotropy ratio \(\lambda_{1}/\lambda_{2}=\mathrm{const}\). Finally, we normalize the fluctuation amplitude by choosing \(\sigma=1\), completing the definition of the fluctuation field \(f\).

\par
\begin{figure*}[htbp]
  \centering
  \includegraphics[width=7cm,height=6cm]{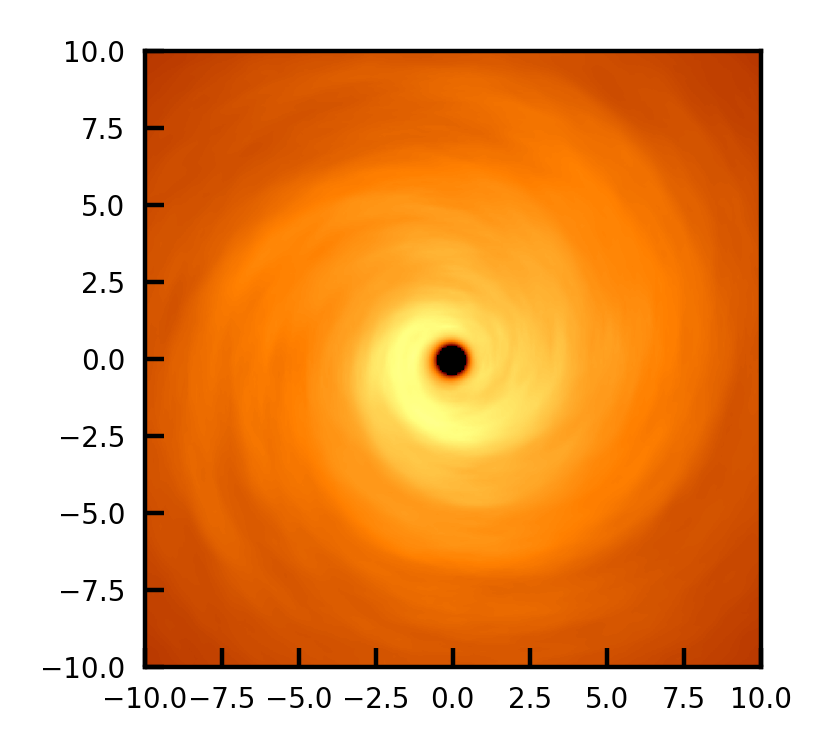}
  \includegraphics[width=7cm,height=6cm]{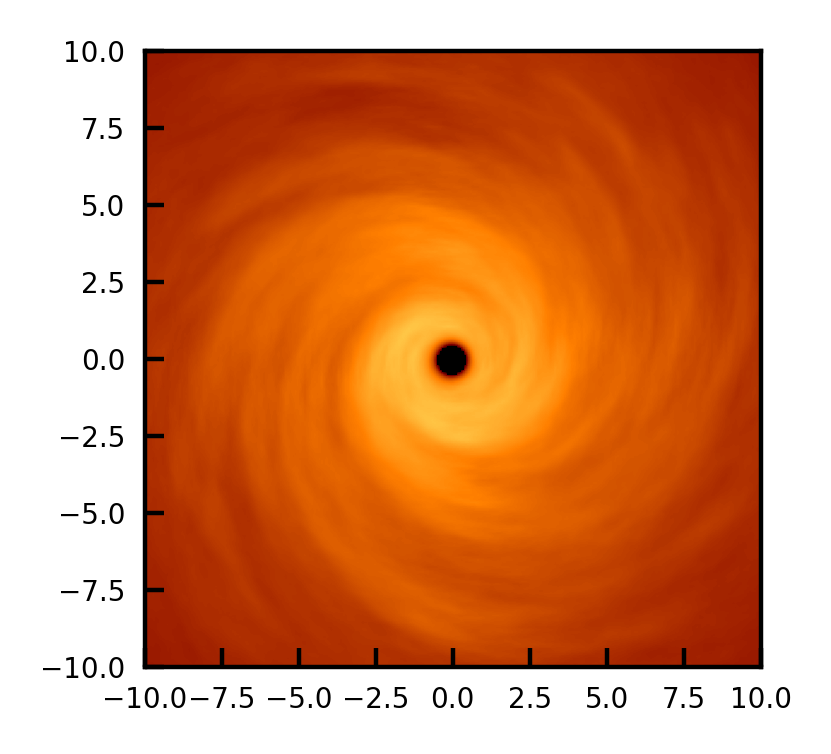}
  \includegraphics[width=7cm,height=6cm]{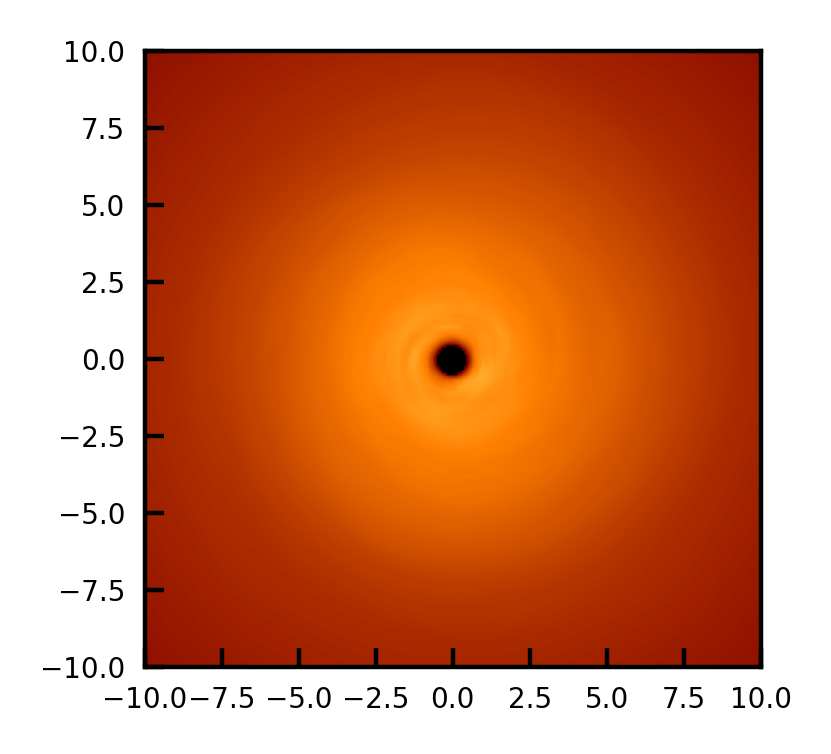}
  \includegraphics[width=7cm,height=6cm]{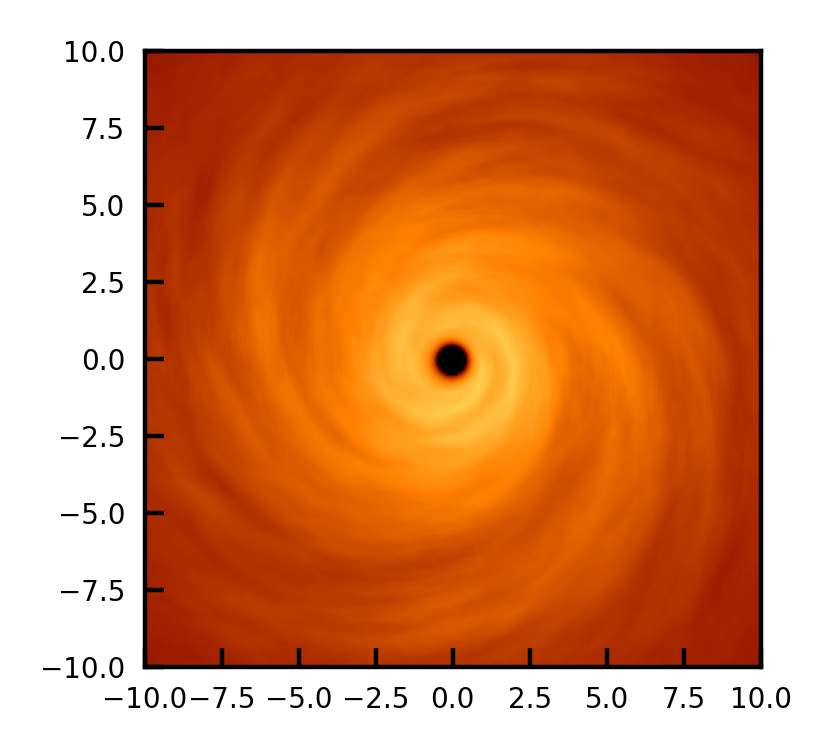}
  \caption {\textcolor{black}{Accretion disk image at $t_{\mathrm{s}} = 20\,M$ based on an inhomogeneous, anisotropic, and time-varying GRF model.}}\label{fig:7}
\end{figure*}

To visualize the dynamical behavior of the accretion disk around a rotating Hayward BH, we construct a time‑evolving brightness map by coupling the fluctuation field \(f\) to the surface brightness \(\mu\). The underlying, time‑averaged radial profile is specified by the envelope function \(g(r)\), where we introduce the dimensionless variable \(x\equiv r_0/r\), with \(r_0\) a characteristic reference radius. We adopt  
\begin{equation}
\label{36}
g(r) = x^{4}e^{-x^{2}},
\end{equation}
which produces a central brightness depression or “shadow,” \textcolor{black}{reminiscent} of the morphology observed in M87$^{*}$, while ensuring an asymptotic \(r^{-4}\) decay at large radii. The instantaneous surface brightness is then modulated by the fractional fluctuations as  
\begin{equation}
\label{37}
\mu(r,\phi,t)
= g(r)\,\exp\!\biggl(\frac{f(r,\phi,t)}{n}\biggr),
\end{equation}
where \(n\) sets the characteristic fluctuation amplitude and thus the effective turbulence strength in the disk atmosphere. We evaluate this time‑variable disk model in a series of numerical experiments, summarized in Fig.~\ref{fig:7}. The fluctuation field is parameterized by \(\lambda_1/r = 5\), an anisotropy ratio \(\lambda_2/\lambda_1 = 0.1\), and a temporal coherence scale \(\lambda_0 = 2\pi/\Omega_K\), where \(\Omega_K = r^{-m}\) denotes the Keplerian angular velocity profile. The first three columns of Fig.~\ref{fig:7} correspond to \(m=1/2\), \(3/2\), and \(2\), respectively. Among these, \(m=3/2\) yields the most \textcolor{black}{pronounced and physically plausible} spiral morphology, closely matching idealized turbulent–accretion flow models. Finally, we contrast \textcolor{black}{prograde} versus \textcolor{black}{retrograde} rotation: the first three panels show prograde disks, while the fourth panel illustrates a retrograde configuration, highlighting the asymmetry in brightness patterns introduced by the BH spin direction.  

\subsection{Temporal Insights into Accretion Flow Variability}
\label{sec:3-3}
\par
To ensure consistency with the kinematic structure of the accretion flow described in Ref.~\cite{29}, we \textcolor{black}{introduce explicit time dependence} following the underlying velocity profile:
\begin{equation}
\label{38}
\vec{v}\equiv \frac{\mathrm{d}r}{\mathrm{d}t}\,\partial_r + \frac{\mathrm{d}\phi}{\mathrm{d}t}\,\partial_\phi
= -\iota\,\partial_r + \Omega\,\partial_\phi,
\end{equation}
where \(\iota\) and \(\Omega\) denote the radial infall rate and azimuthal angular velocity, respectively. In Cartesian coordinates, this \textcolor{black}{yields}
\begin{equation}
\label{39}
v_x(\mathbf{x}_{\rm s}) = -\frac{x_{\rm s}}{r_{\rm s}}\,\iota - y_{\rm s}\,\Omega, 
\qquad 
v_y(\mathbf{x}_{\rm s}) = -\frac{y_{\rm s}}{r_{\rm s}}\,\iota + x_{\rm s}\,\Omega,
\end{equation}
where \(\mathbf{x}_{\rm s}=(x_{\rm s},y_{\rm s})\) and \(r_{\rm s}=\sqrt{x_{\rm s}^2+y_{\rm s}^2}\). \textcolor{black}{To align spatial correlations with the flow geometry}, we prescribe the orientation angle
\begin{equation}
\label{40}
\theta(\mathbf{x}_{\rm s}) = \arctan2(y_{\rm s},-x_{\rm s}) + \theta_{\angle},
\end{equation}
which \textcolor{black}{sets} the principal axis \(\mathbf{u}_1(\mathbf{x}_{\rm s})\) in the equatorial plane at a pitch angle \(\theta_{\angle}\) relative to the radial direction. Consequently, the velocity-aligned structure \textcolor{black}{exhibits} a spiral morphology characterized by the opening angle \(\theta_{\angle}\). We \textcolor{black}{adopt} \(\theta_{\angle}\approx20^\circ\) to match the spiral pitch angles observed in GRMHD simulations~\cite{39,40}, thereby ensuring \textcolor{black}{consistency} between the statistical emission model and the underlying fluid dynamics.

\par
After constructing the time‑dependent GRF for the accretion disk, we adopt the \textcolor{black}{fast‑light} approximation to perform ray tracing in the spacetime of a rotating Hayward BH. In this framework, photons emitted from the disk at different emission times \(t_{\rm s}^{(n)}(\alpha,\beta)\) but arriving at the same image‑plane coordinate \((\alpha,\beta)\) are assumed to be registered \textcolor{black}{simultaneously} in the observer’s frame. In reality, each photon traverses a distinct geodesic and accumulates a \textcolor{black}{gravitational and geometric} time delay \(\Delta t = t_{\rm o} - t_{\rm s}^{(n)}\). The fast‑light approximation \textcolor{black}{neglects these delays}, treating all photons at a given observer time as if they originated from a single, frozen emission snapshot. Under this assumption, the image‑plane intensity reduces to
\begin{equation}
\label{41}
I_{\rm o}(\alpha,\beta)
=\sum_{n=0}^{N(\alpha,\beta)}\zeta_n\,g^3\bigl(r_{\rm s}^{(n)},\alpha,\beta\bigr)\,
I_{\rm s}\bigl(r_{\rm s}^{(n)},\phi_{\rm s}^{(n)}\bigr),
\end{equation}
where \(g\) is the redshift factor, \(\zeta_n\) are the geodesic weighting coefficients, and \(I_{\rm s}\) is the equatorial source brightness. Equivalently, one \textcolor{black}{holds} the emission time \(t_{\rm s}\) fixed while allowing the observed time \(t_{\rm o}\) to vary across \((\alpha,\beta)\). For stationary or slowly evolving accretion flows, this approximation remains \textcolor{black}{both accurate and computationally efficient}, effectively treating each \textmd{INOISY} realization as a static emission map. Consequently, the fast‑light assumption is well suited for modeling radiative transfer in quasi‑steady disk configurations.

\par
In general‑relativistic radiative transfer, photons emitted from different regions of the accretion disk follow distinct geodesics and therefore accrue different propagation delays to the observer. This \textcolor{black}{slow‑light} effect is particularly critical for rapidly evolving structures such as orbiting hotspots or propagating spiral waves. The time delay for a photon arriving at image‑plane coordinate \((\alpha,\beta)\) is given by:
\begin{equation}
\label{42}
\Delta t(\alpha,\beta)
= t_{\rm o} - t_{\rm s}^{(n)}(\alpha,\beta),
\end{equation}
where \(t_{\rm o}\) is the observer time and \(t_{\rm s}^{(n)}\) denotes the coordinate time of emission along the \(n\)th null geodesic. A full slow‑light treatment \textcolor{black}{requires computing these delays along each geodesic and reconstructing the observed image from multiple emission snapshots}. While this approach \textcolor{black}{provides maximal physical fidelity for highly dynamic scenarios}, it is \textcolor{black}{computationally prohibitive} for time‑dependent GRF models with coupled spatial and temporal fluctuations.

\par
\begin{figure*}[htbp]
  \centering
  \includegraphics[width=5cm,height=5cm]{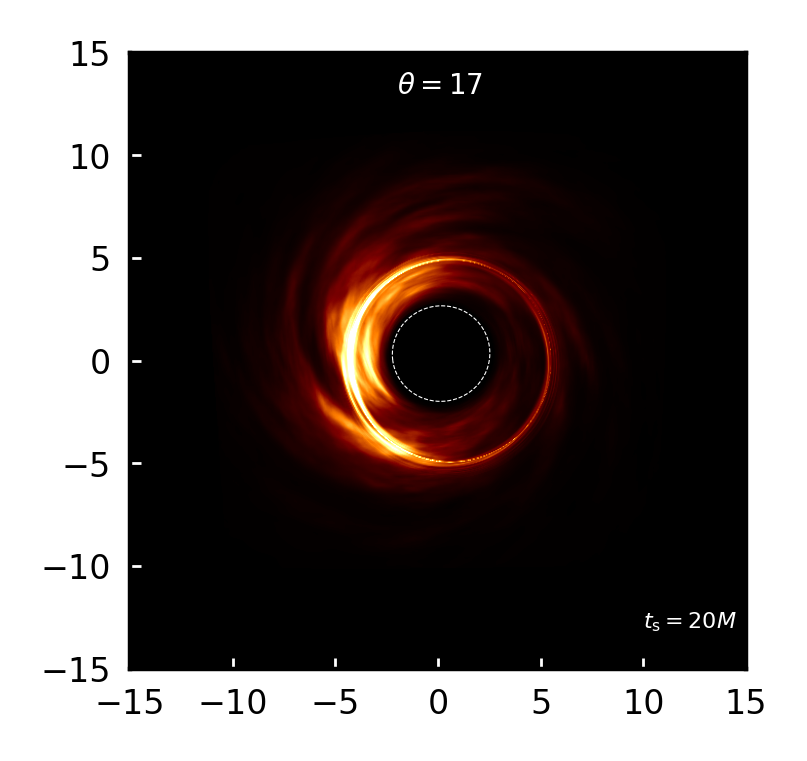}
  \includegraphics[width=5cm,height=5cm]{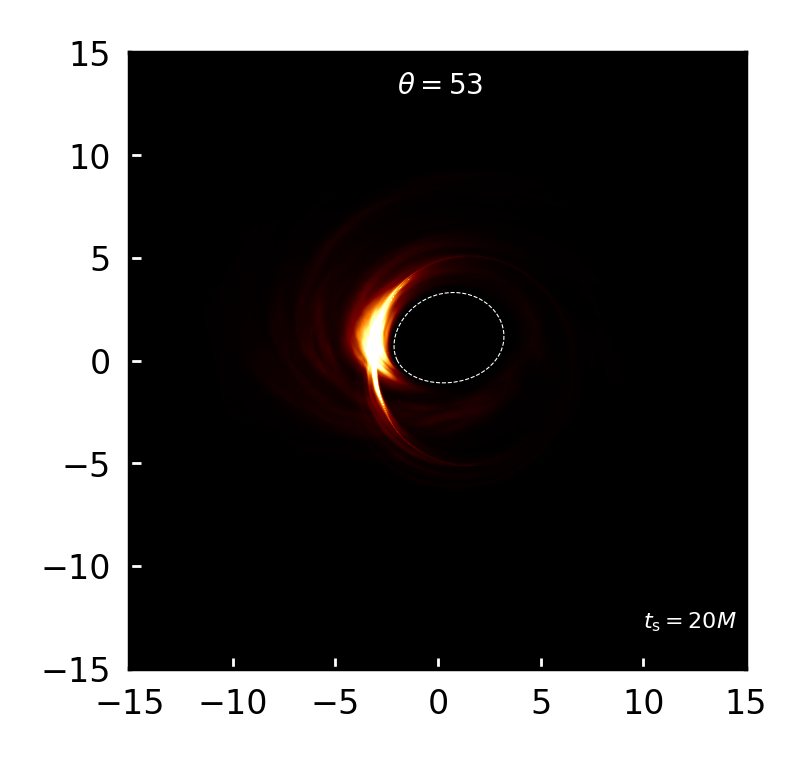}
  \includegraphics[width=5cm,height=5cm]{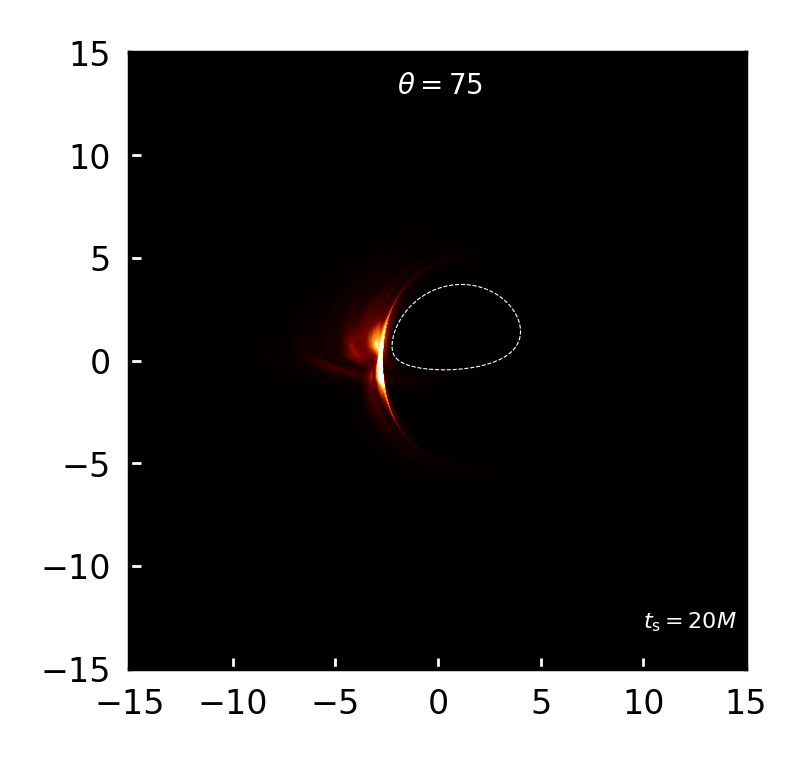}
  \caption {\textcolor{black}{Fast-light images of a rotating Hayward BH at \(t_{\mathrm{s}} = 20M\) as seen by observers at different inclination angles. \textbf{Left panel}: viewing angle \( \theta_0 = 17^\circ \); \textbf{middle panel}: \( \theta_0 = 53^\circ \); \textbf{right panel}: \( \theta_0 = 75^\circ \). The BH mass is $M=1$ and magnetic charge is $g=0.5$. The observed asymmetry and brightness enhancement in the upper hemisphere increase with inclination, reflecting relativistic Doppler beaming effects in a prograde accretion flow.
}}\label{fig:8}
\end{figure*}

In this work, we \textcolor{black}{employ} the \textit{fast‑light approximation}, which assumes that all photons reaching the observer at a given time originate from a single, frozen emission snapshot \(t_{\rm s}=\mathrm{const}\). This simplification \textcolor{black}{has been shown} to hold for stationary or slowly varying accretion flows and \textcolor{black}{permits} efficient rendering of synthetic images without explicitly tracking individual light‑travel delays. \textcolor{black}{Synthetic fast‑light images} computed for various viewing inclinations are presented in Fig.~\ref{fig:8}. As the observer’s inclination angle increases, \textcolor{black}{a pronounced brightness enhancement} emerges in the northern hemisphere of the image, consistent with the Doppler‑boosting signature of a prograde accretion disk. \textcolor{black}{Specifically}, the left panel of Fig.~\ref{fig:8} corresponds to an inclination of \(17^\circ\). \textcolor{black}{Shadowing and scattering‑induced asymmetries and structural distortions become increasingly evident}, in close agreement with predictions from GRMHD simulations.

\par
As the observer inclination increases to \(53^\circ\) and \(75^\circ\), the image progressively concentrates brightness in the northwest quadrant, with the inner shadow boundary traced by the white dashed line. In this analysis, we consider a rotating Hayward BH with dimensionless spin parameter \(a=0.8\) and magnetic charge \(g=0.5\). Variations in these parameters are deferred to future work, as our primary focus here is on the temporal evolution of the BH image coupled to the surrounding accretion flow. The resulting image morphology aligns closely with single‑frame outputs from high‑resolution GRMHD simulations, thereby validating the physical plausibility of our stochastic, time‑dependent model. Fast‑light ray tracing enables reconstruction of the disk morphology at discrete time intervals by integrating the time‑variable emission structure with the curved spacetime of the rotating Hayward BH. This approach parallels imaging techniques applied to $M87^{*}$ and $Sgr A^{*}$, where multiple temporally spaced snapshots are synthesized to capture intrinsic variability. These snapshots are subsequently averaged to yield a time-integrated image, providing a comprehensive representation of the long‑term dynamical behavior of the accretion flow.

\par
\begin{figure*}[htbp]
  \centering
  \includegraphics[width=7cm,height=5cm]{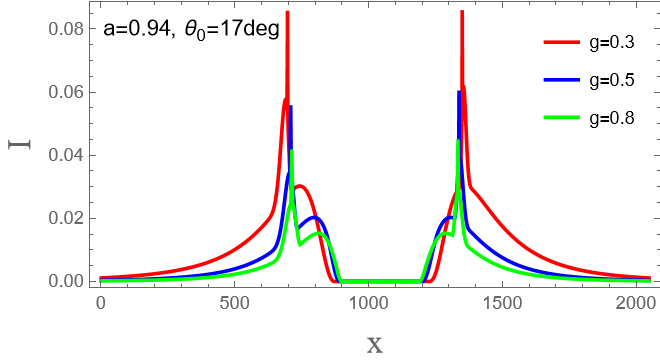}
  \includegraphics[width=7cm,height=5cm]{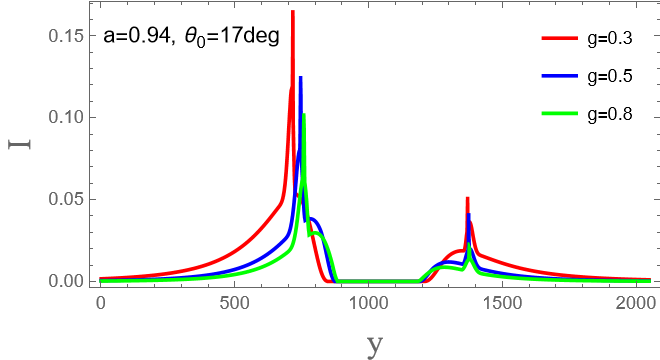}
  \caption {\textcolor{black}{Luminosity profiles of Hayward BH for various magnetic charge parameters. The left panel shows intensity distributions along the \( x \)-axis, and the right panel displays those along the \( y \)-axis.}}\label{fig:81}
\end{figure*}
\begin{figure*}[htbp]
  \centering
  \includegraphics[width=5cm,height=4.8cm]{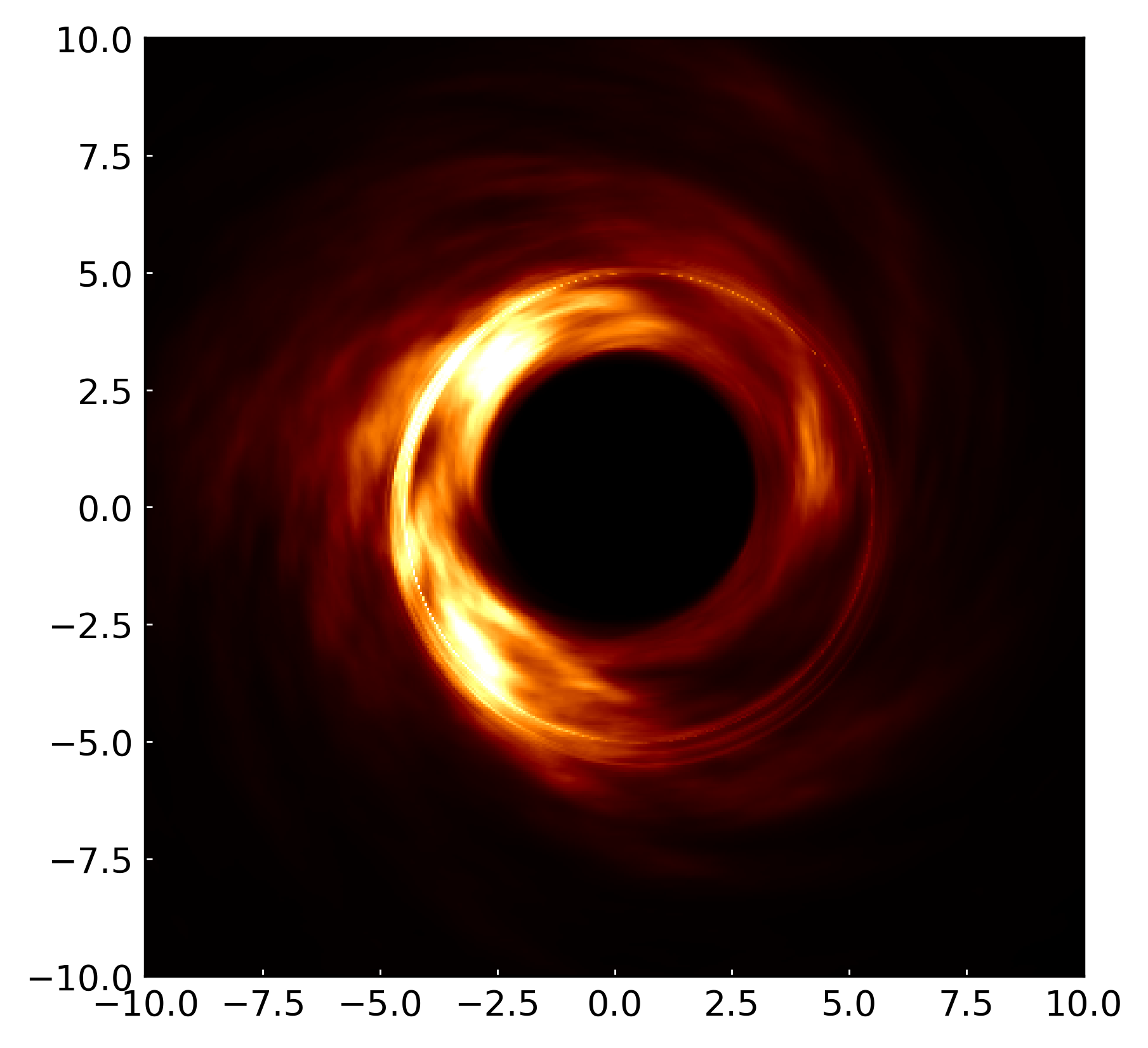}
  \includegraphics[width=5cm,height=4.8cm]{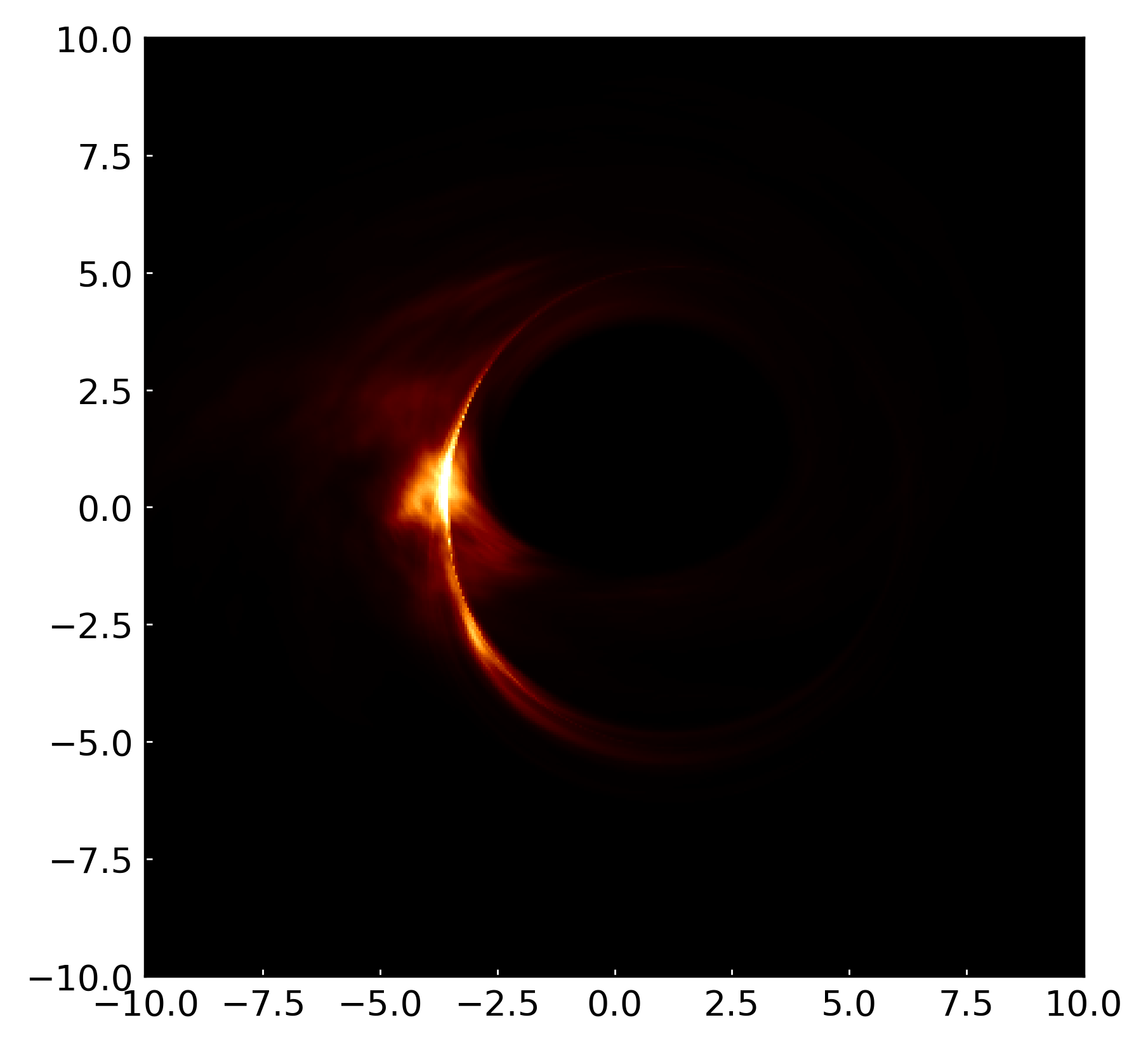}
  \includegraphics[width=5cm,height=4.8cm]{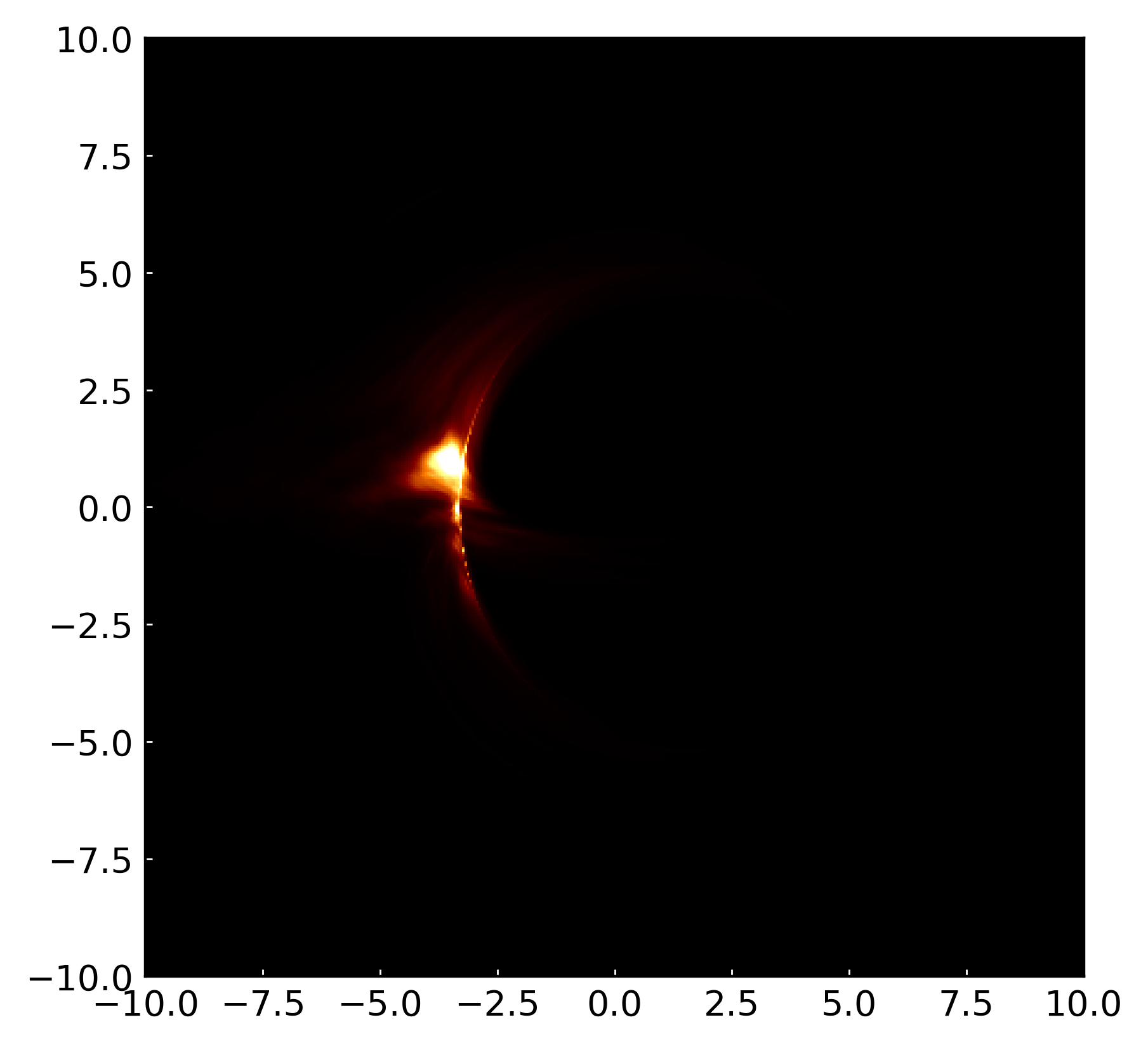}
  \caption {\textcolor{black}{The same as Fig. 8 but for $g=0.8$.
}}\label{fig:82}
\end{figure*}

\textcolor{black}{In addition, we also investigate the influence of varying magnetic charges on the images. As illustrated in Fig.~\ref{fig:81}, the luminosity profiles along the \( x \)-axis and \( y \)-axis differ significantly. Furthermore, the luminosity gradually decreases with increasing magnetic charge of the BH. Although the variation in magnetic charge itself does not directly alter the radiation mechanism, a stronger magnetic field can disperse accreting material, reducing the accretion rate and consequently lowering the luminosity. Alternatively, strong magnetic fields may also facilitate energy release through non-radiative channels, resulting in diminished brightness. Fig.~\ref{fig:82} presents the simulation results corresponding to increasing magnetic charge.}

\par
We \textcolor{black}{augment} the conventional ray‑tracing framework to incorporate the \textit{slow‑light} effect, thereby capturing subtle variations in photon emission times across the image plane. This \textcolor{black}{enhanced} formulation provides a more accurate treatment of relativistic photon propagation by explicitly accounting for time delays arising from both emission and scattering processes in the immediate vicinity of the BH. By employing the formalism of Eq.~(\ref{41}), we obtain results that fully \textcolor{black}{resolve} the temporal evolution of the accretion flow, offering a dynamically time‑resolved perspective of the system. This methodology not only \textcolor{black}{improves} the temporal fidelity of the simulated images but also \textcolor{black}{ensures} a self‑consistent coupling between the evolving spacetime geometry, the time‑dependent disk emission, and the photon geodesics.

\par
Under the \textit{fast‑light approximation}, \textcolor{black}{photons arriving at a single observer time} are assumed to emanate simultaneously from a \textcolor{black}{frozen emission hypersurface} \(t_{\rm s}=\mathrm{const}\), \textcolor{black}{thereby producing} a \textcolor{black}{time‑independent} brightness map.
\[
I_{\rm fast}(\boldsymbol{x}) = I\big(r_{\rm s}^{(n)}(\boldsymbol{x}), \phi_{\rm s}^{(n)}(\boldsymbol{x}), t_{\rm s}\big),
\]
and corresponding visibility function
\begin{equation}
\label{43}
V_{\rm fast}(\boldsymbol{u}) = \iint I_{\rm fast}(\boldsymbol{x})\, e^{-2\pi i\, \boldsymbol{u} \cdot \boldsymbol{x}}\, d^2\boldsymbol{x}.
\end{equation}
In contrast, the slow-light model accounts for trajectory-dependent emission times \( t_{\rm s}^{(n)}(\boldsymbol{x}) \), leading to
\[
I_{\rm slow}(\boldsymbol{x}) = I\big(r_{\rm s}^{(n)}(\boldsymbol{x}), \phi_{\rm s}^{(n)}(\boldsymbol{x}), t_{\rm s}^{(n)}(\boldsymbol{x})\big),
\]
and its corresponding visibility function
\begin{equation}
\label{44}
V_{\rm slow}(\boldsymbol{u}) = \iint I_{\rm slow}(\boldsymbol{x})\, e^{-2\pi i\, \boldsymbol{u} \cdot \boldsymbol{x}}\, d^2\boldsymbol{x}.
\end{equation}
The residual between the two approaches can be quantified as
\begin{equation}
\label{45}
\delta V(\boldsymbol{u}) = V_{\rm slow}(\boldsymbol{u}) - V_{\rm fast}(\boldsymbol{u}),
\end{equation}
providing a diagnostic of the phase and amplitude errors introduced by neglecting relativistic time delays in the fast-light approximation.

\par
\begin{figure*}[htbp]
  \centering
  \includegraphics[width=5cm,height=4.8cm]{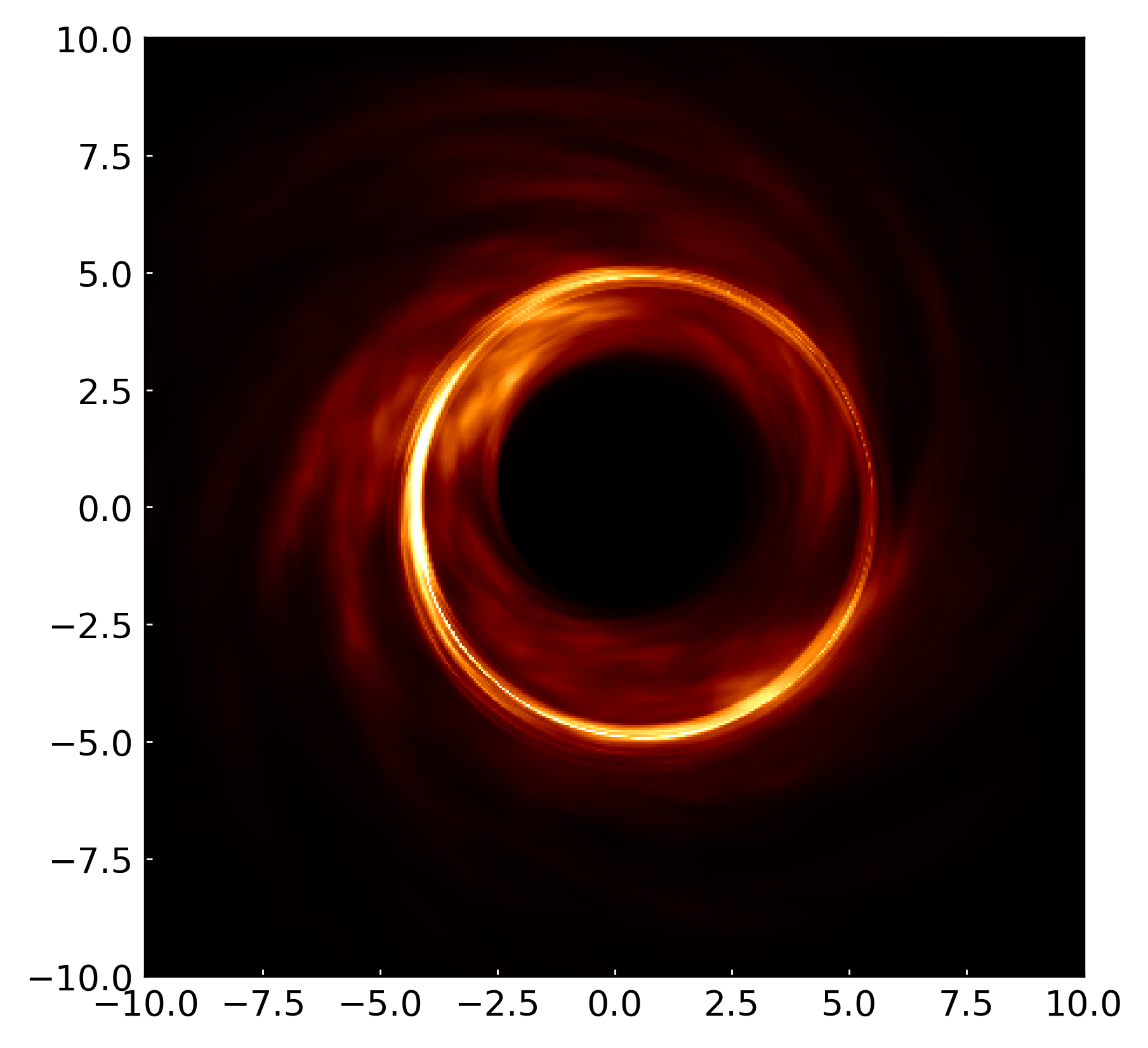}
  \includegraphics[width=5cm,height=4.8cm]{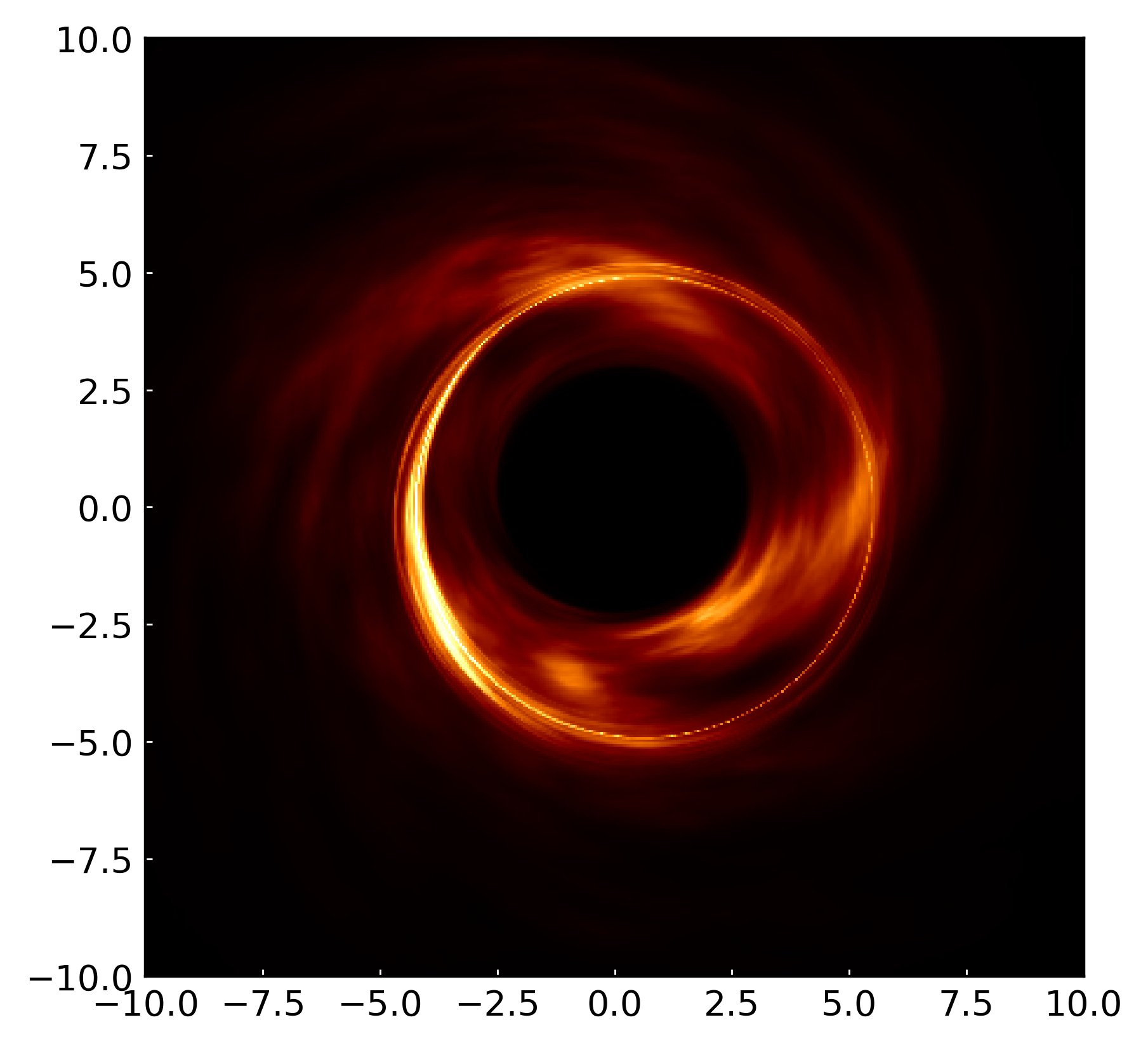}
  \includegraphics[width=5cm,height=4.8cm]{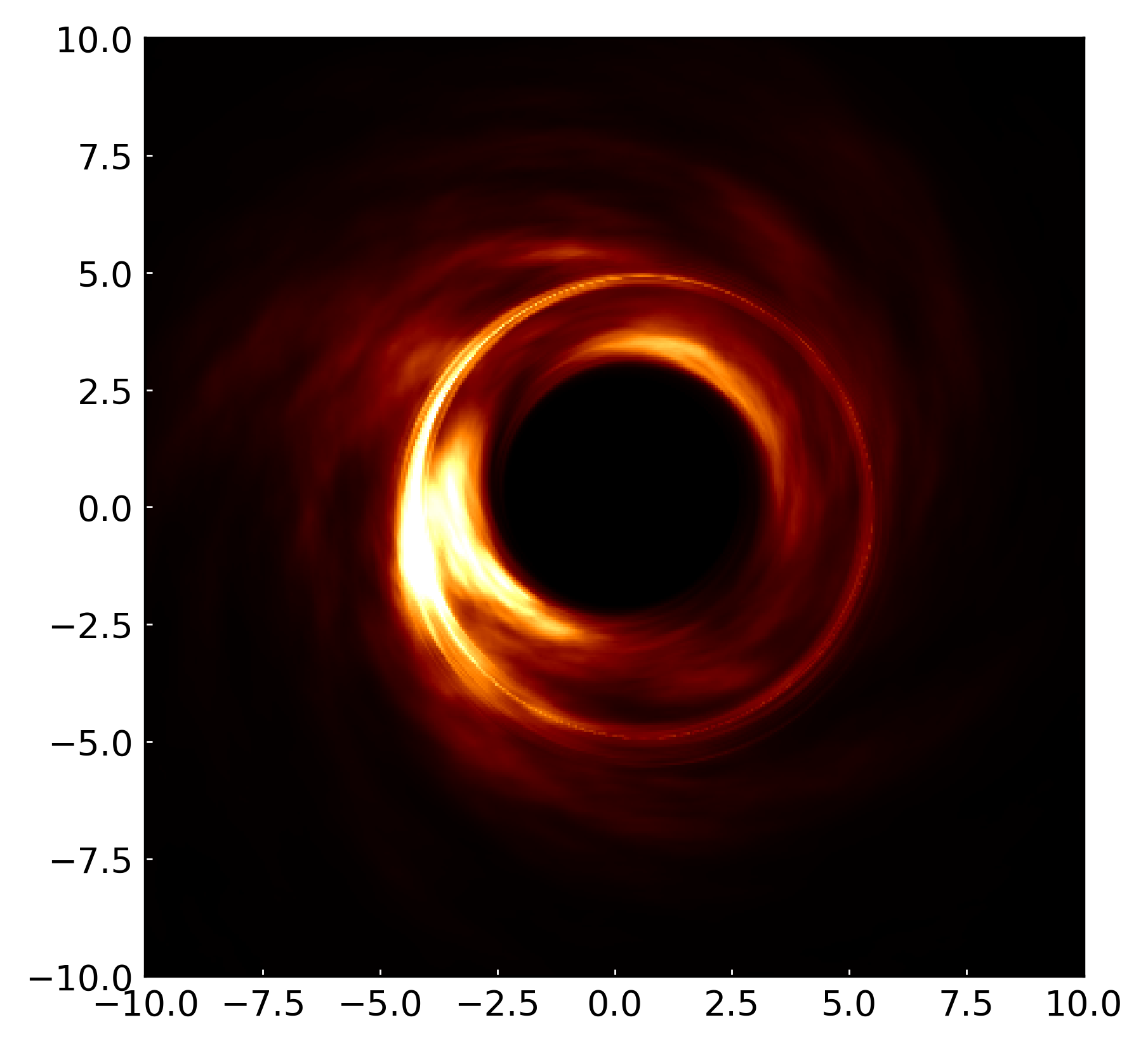}
  \caption {\textcolor{black}{Slow-light images of a rotating Hayward BH at an observer inclination angle of \( \theta_0 = 17^\circ \). The snapshots are taken at intervals of \(100\,M\) within a total time span of \(300\,M\). Under the slow-light approximation, the brightness distributions illustrate the relativistic Doppler boosting and time-delay effects associated with photon trajectories.}}\label{fig:9-1}
\end{figure*}
\begin{figure*}[htbp]
  \centering
  \includegraphics[width=5cm,height=4.8cm]{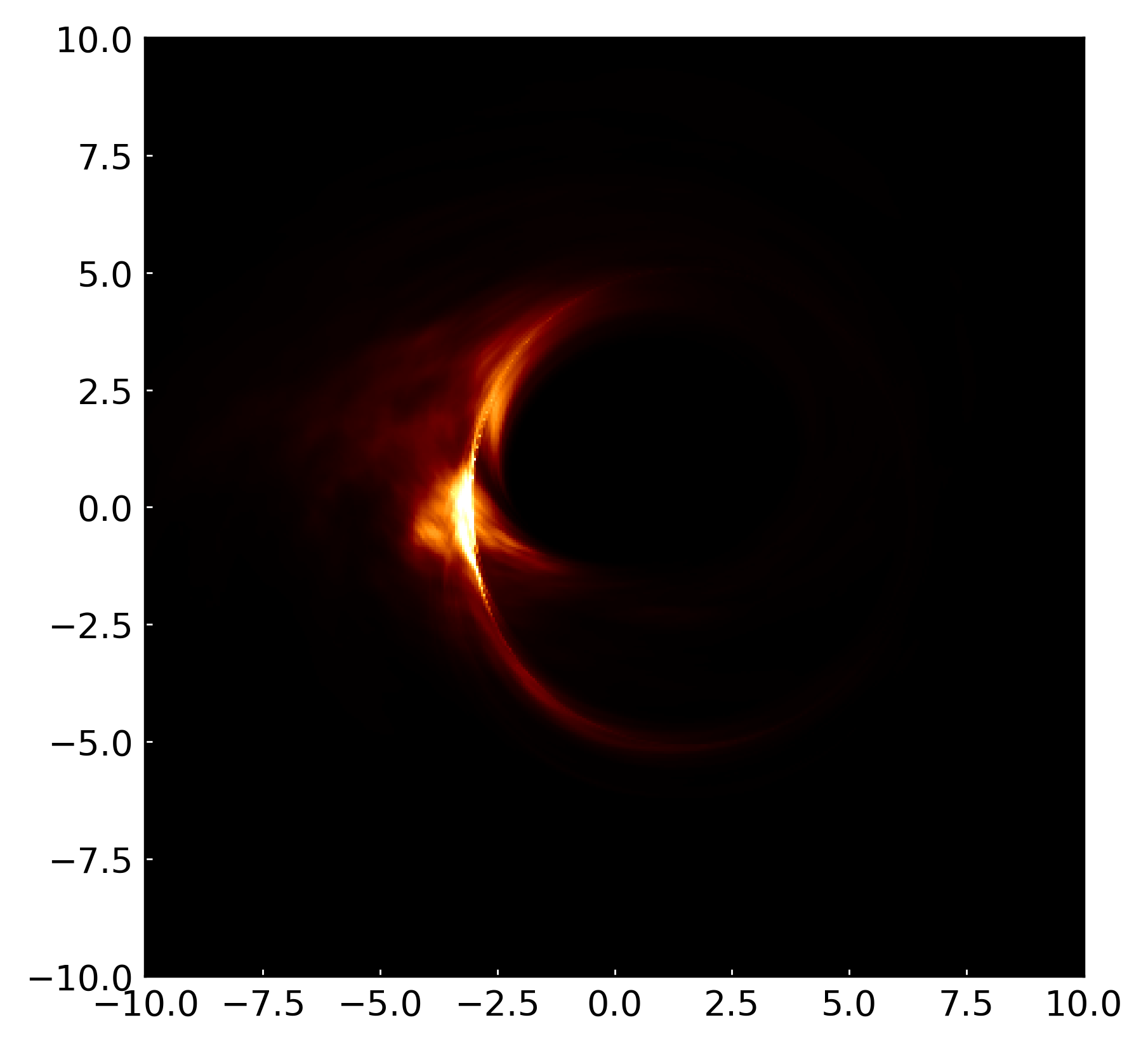}
  \includegraphics[width=5cm,height=4.8cm]{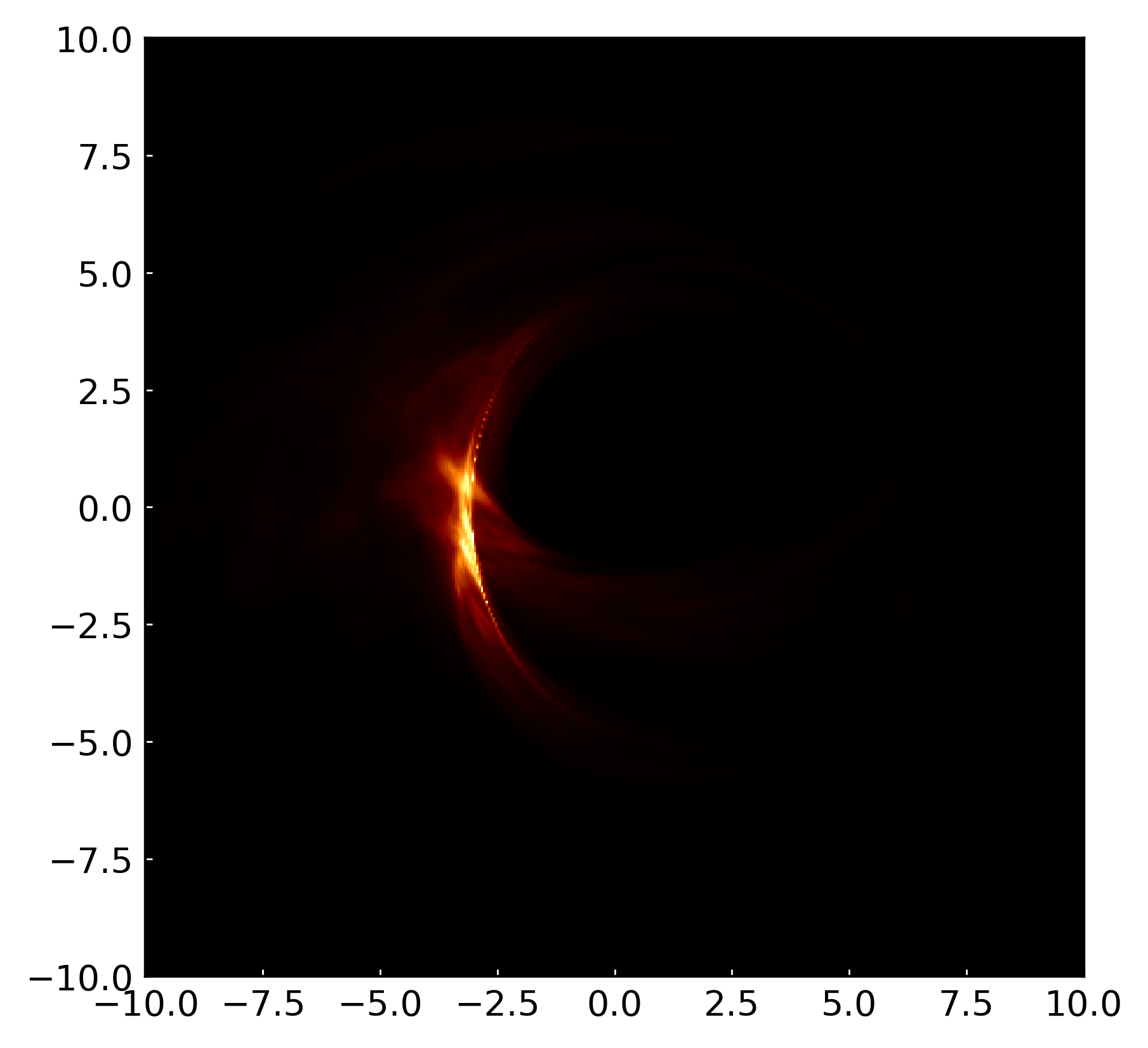}
  \includegraphics[width=5cm,height=4.8cm]{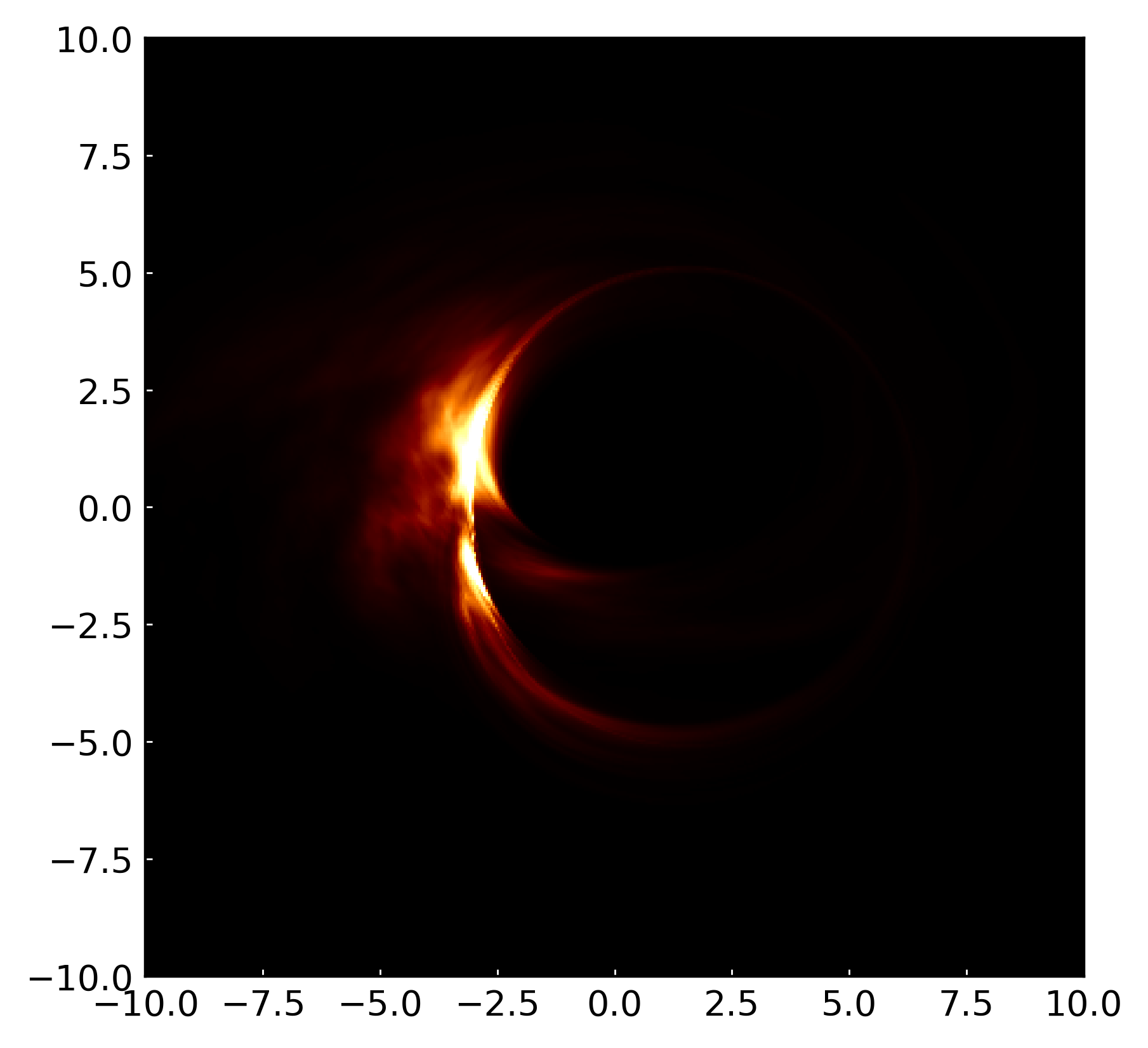}
  \caption {\textcolor{black}{The same as Fig. 11 but for \( \theta_0 = 53^\circ \).}}\label{fig:9-2}
\end{figure*}
\begin{figure*}[htbp]
  \centering
  \includegraphics[width=5cm,height=4.8cm]{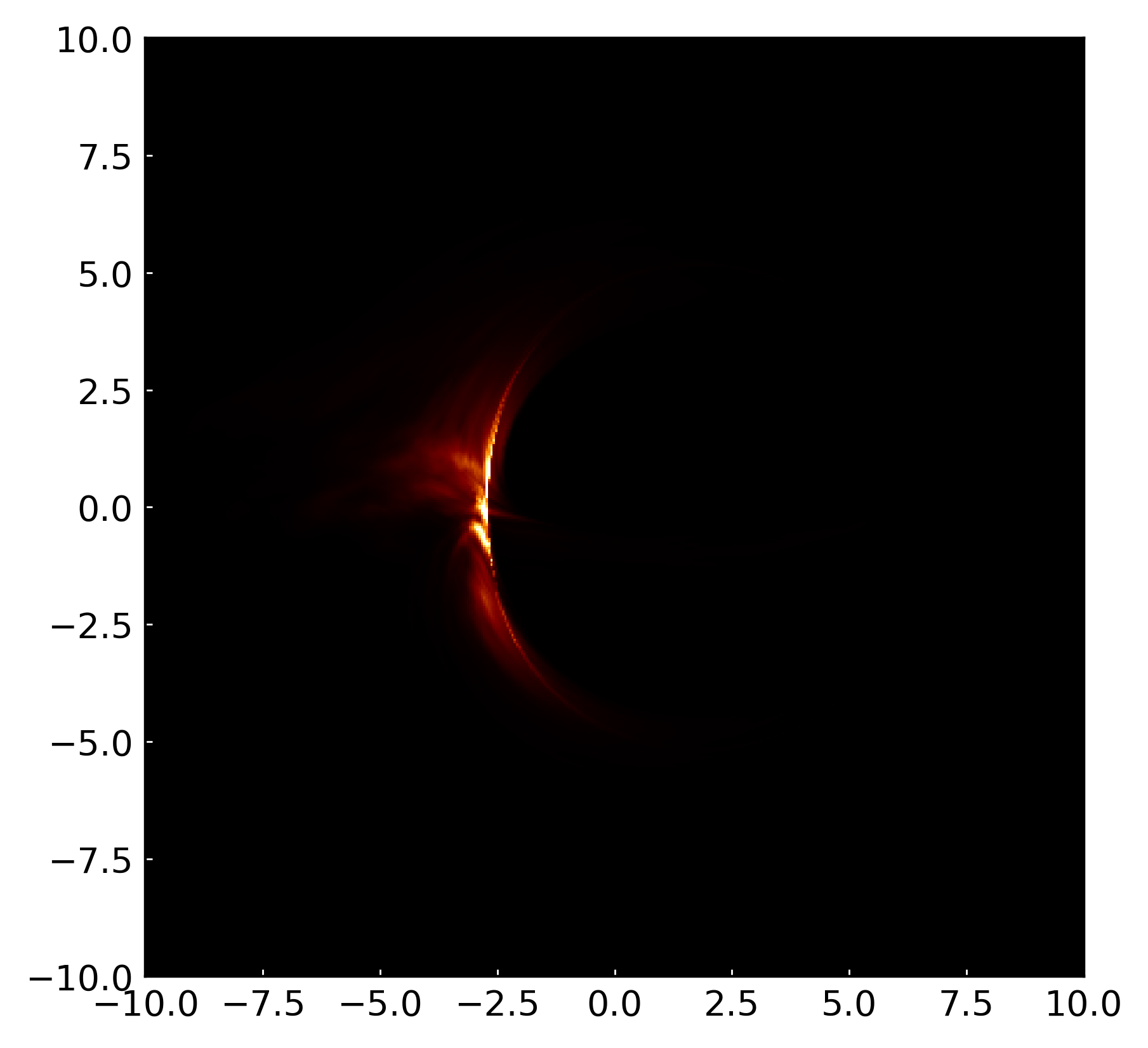}
  \includegraphics[width=5cm,height=4.8cm]{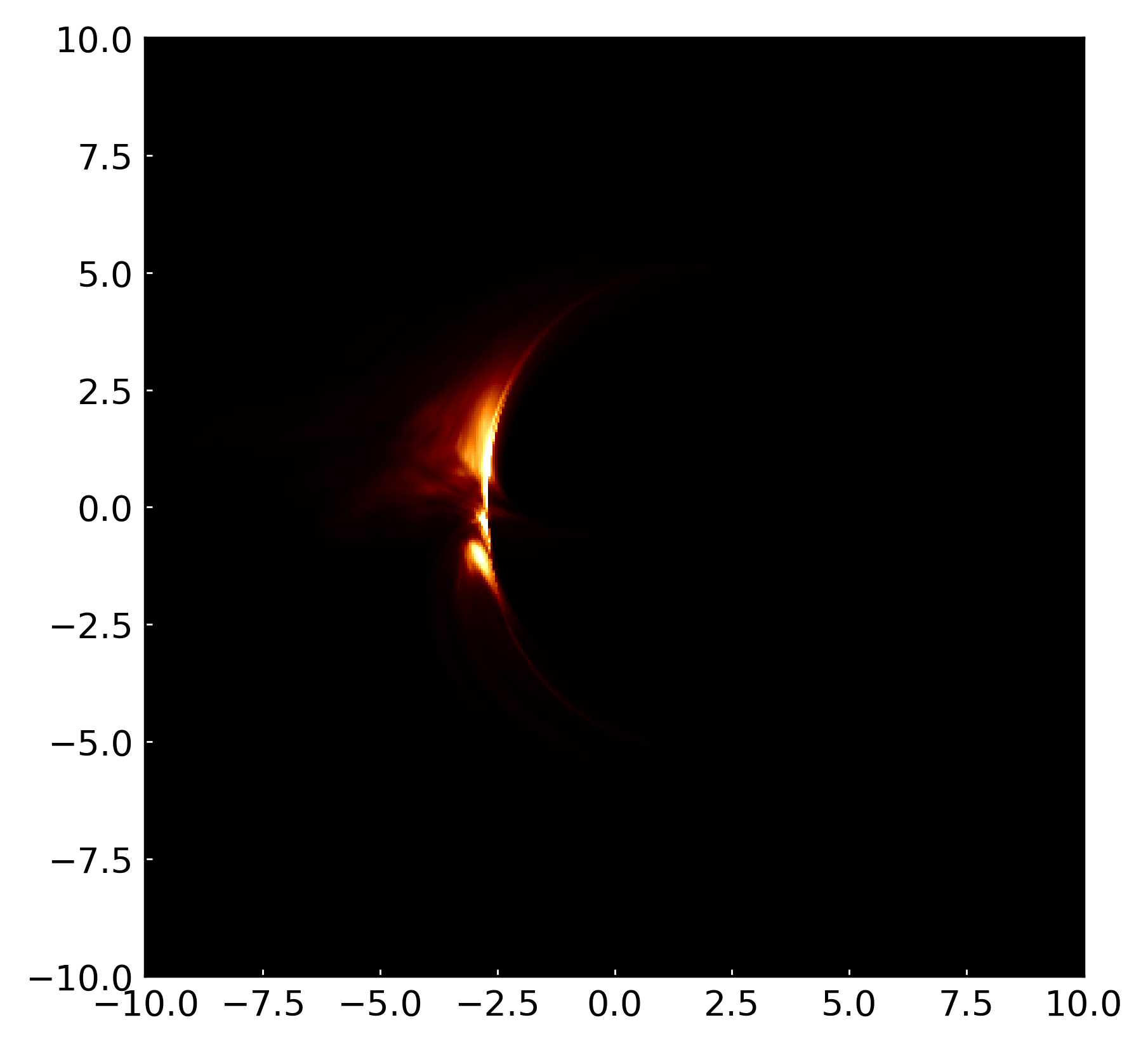}
  \includegraphics[width=5cm,height=4.8cm]{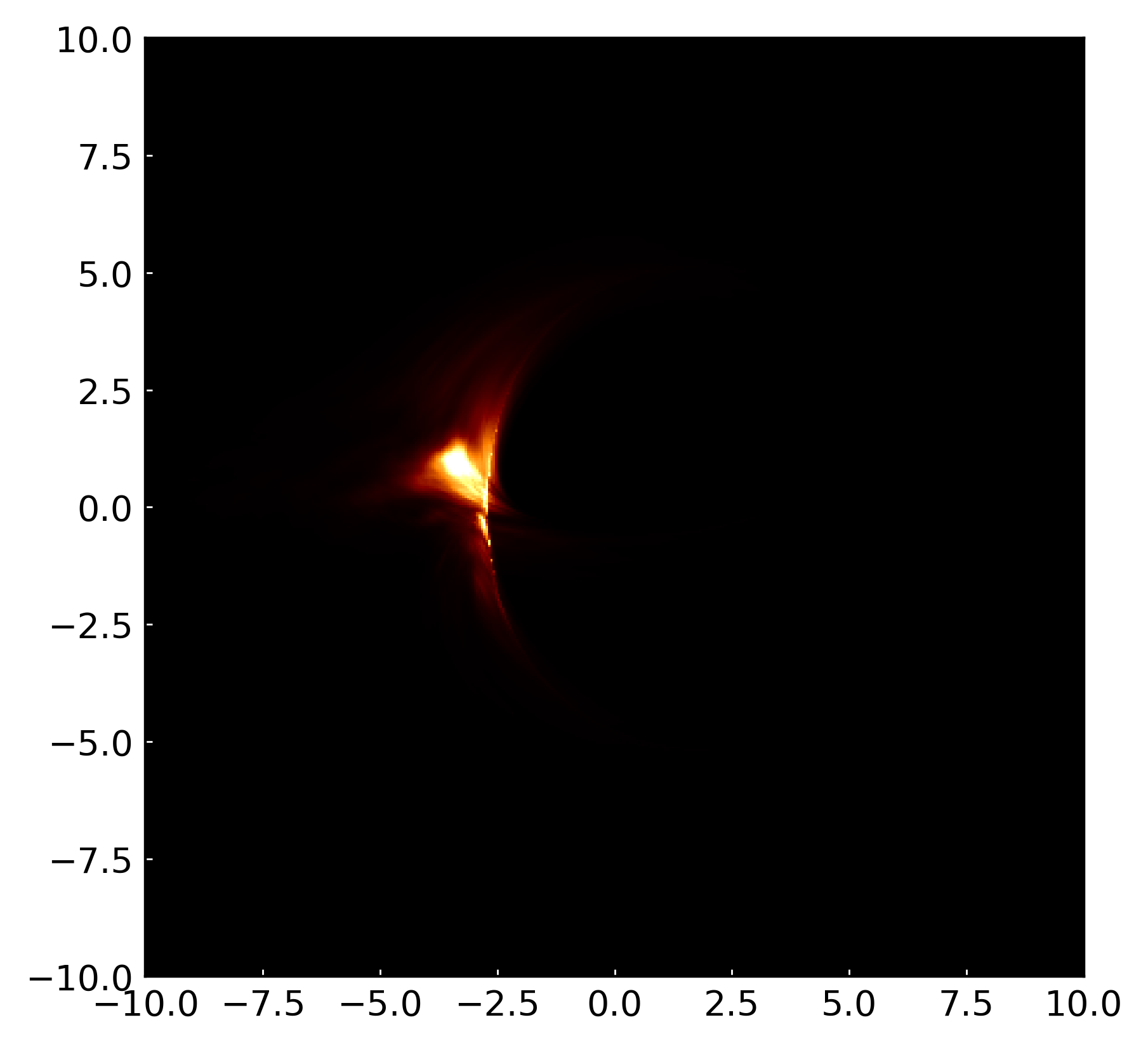}
  \caption {\textcolor{black}{The same as Fig. 11 but for \( \theta_0 = 75^\circ \).}}\label{fig:9-3}
\end{figure*}

The \textcolor{black}{slow‑light} ray‑tracing approach introduces \textcolor{black}{significant} complexity into the imaging of accretion flows by incorporating trajectory‑dependent emission times. In this \textcolor{black}{formalism}, the image slice associated with each image‑plane coordinate \((\alpha,\beta)\) corresponds to an \textcolor{black}{extended} emission window, denoted \(t_{\rm s}^{(n)}(\alpha,\beta)\). Therefore, to \textcolor{black}{construct} a time‑lapse sequence with observational duration \(T_{\rm o}\), it is \textcolor{black}{necessary} to simulate the accretion flow over an intrinsic time interval \(T_{\rm s}\gtrsim T_{\rm o}\). In Figs.~\ref{fig:9-1}--\ref{fig:9-3}, we present snapshots of a rotating Hayward BH captured at intervals of \(40\,M\). As in the fast‑light scenario shown in Fig.~\ref{fig:8}, each frame exhibits \textcolor{black}{transient bright} features \textcolor{black}{closely matching} those seen in high‑resolution GRMHD simulations. Notably, these features \textcolor{black}{correspond} to those in Fig.~1 of Ref.~\cite{12}, which was generated by averaging 100 equally spaced snapshots over a total duration of \(1000\,M\).
\begin{figure*}[htbp]
  \centering
  \includegraphics[width=17cm,height=10cm]{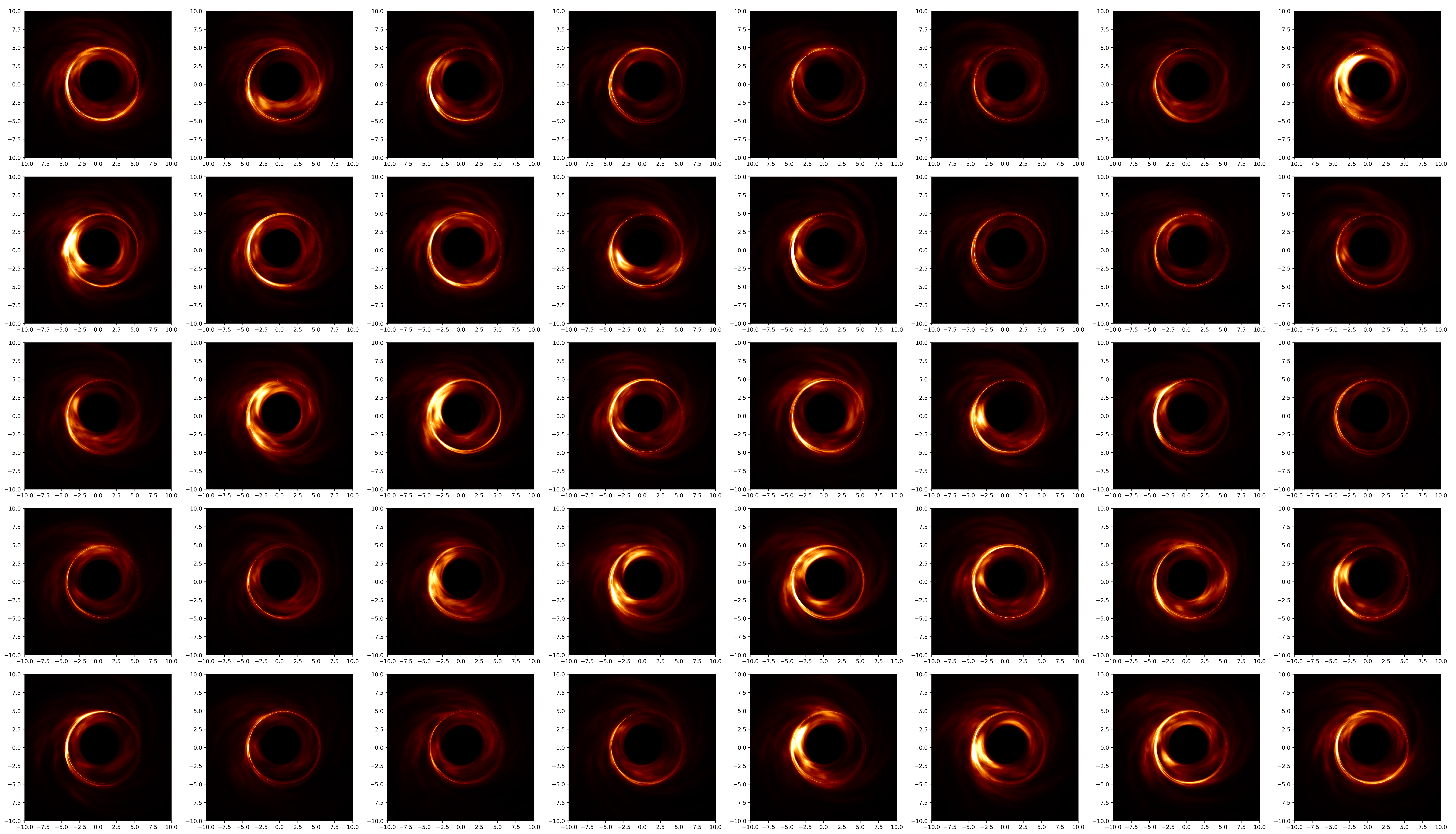}
  \caption {\textcolor{black}{A sequence of snapshots simulating the rotating Hayward BH over a total duration of 400 seconds, with a time interval of 10 seconds between each frame. The observer's inclination angle is set to \( \theta_0 = 17^\circ \).}}\label{fig:10}
\end{figure*}
\begin{figure*}[htbp]
  \centering
  \includegraphics[width=17cm,height=10cm]{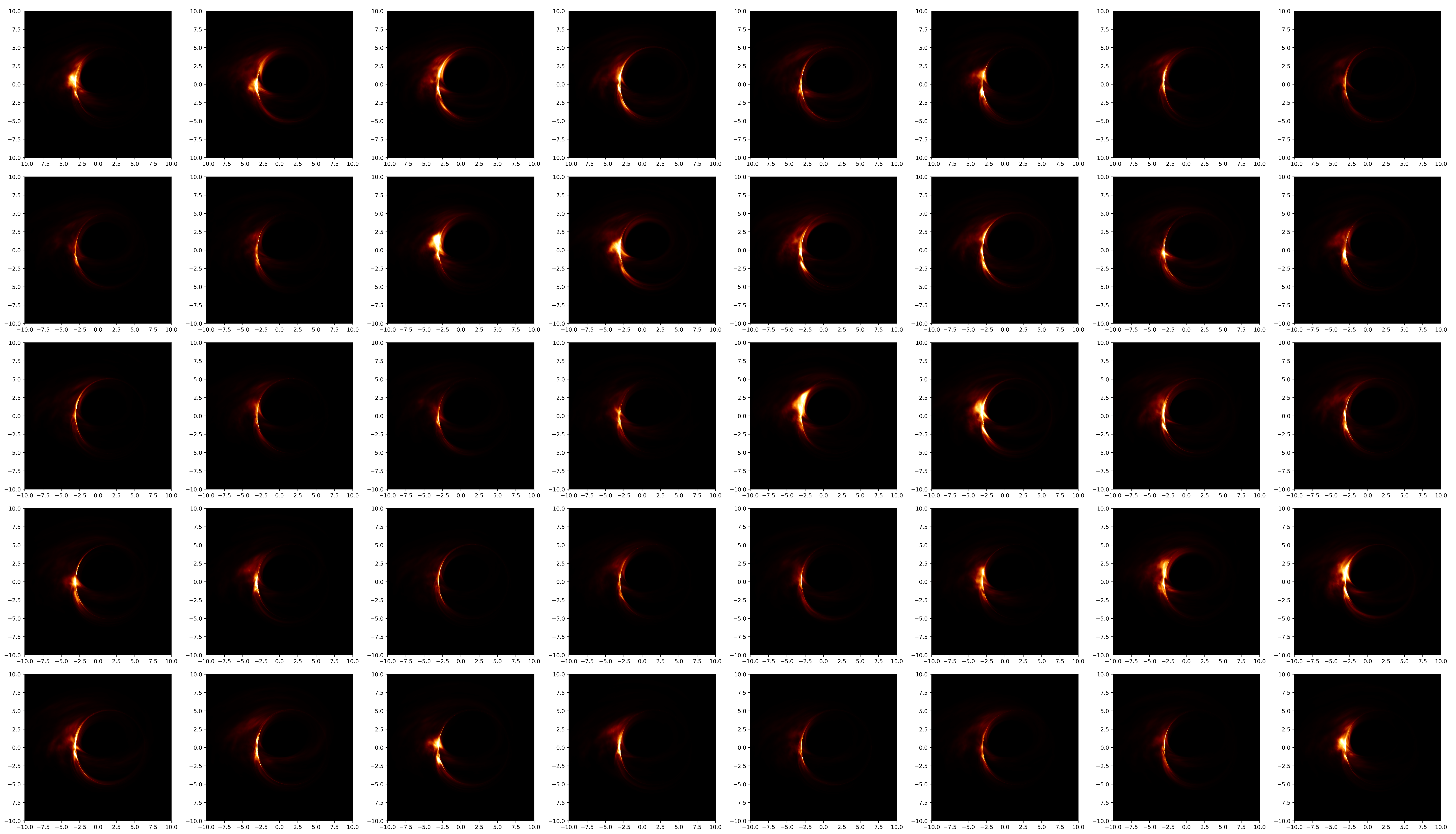}
  \caption {\textcolor{black}{A sequence of snapshots simulating the rotating Hayward BH over a total duration of 400 seconds, with a time interval of 10 seconds between each frame. The observer's inclination angle is set to \( \theta_0 = 53^\circ \).}}\label{fig:11}
\end{figure*}
\begin{figure*}[htbp]
  \centering
  \includegraphics[width=17cm,height=10cm]{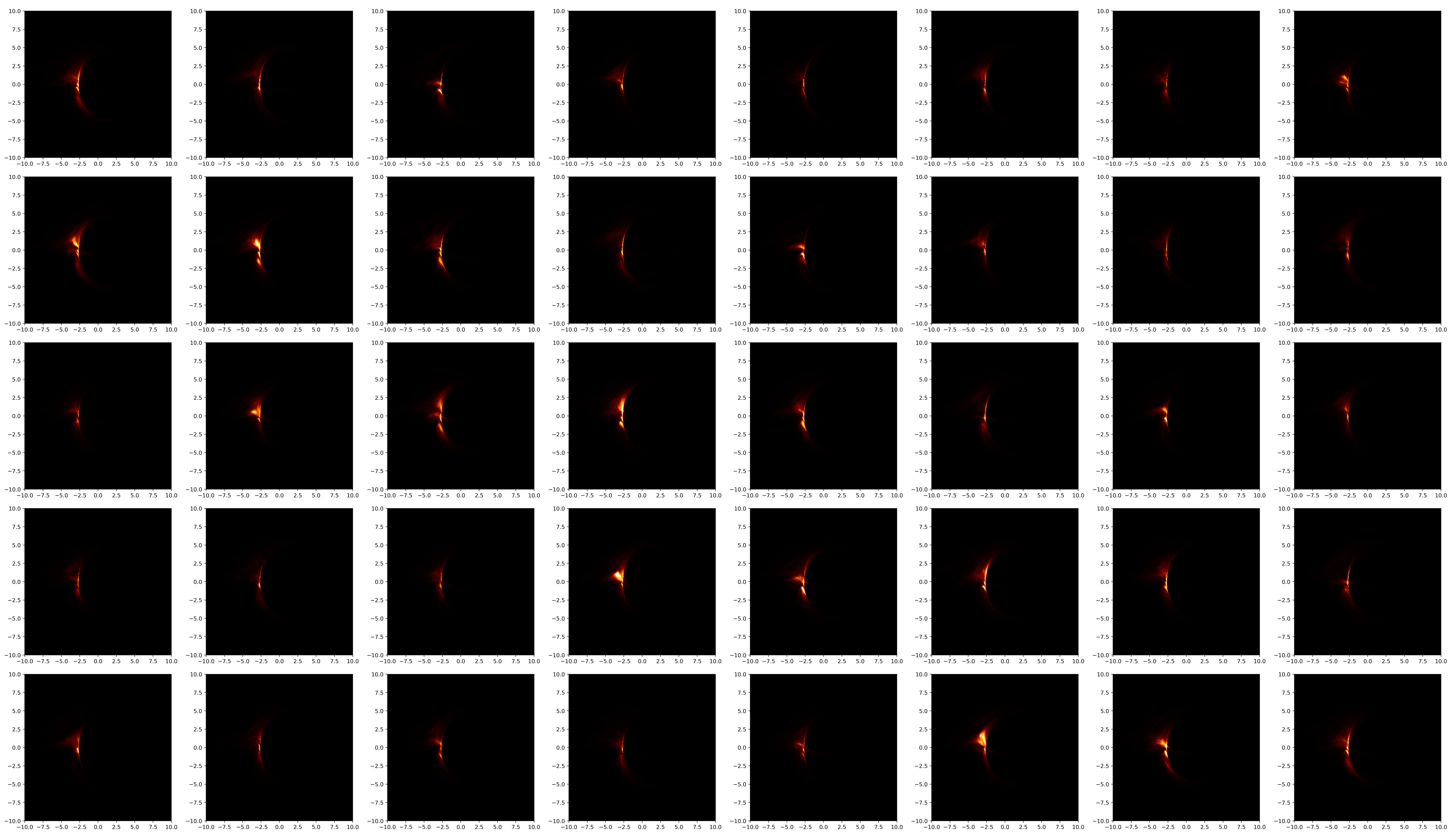}
  \caption {\textcolor{black}{A sequence of snapshots simulating the rotating Hayward BH over a total duration of 400 seconds, with a time interval of 10 seconds between each frame. The observer's inclination angle is set to \( \theta_0 = 75^\circ \).}}\label{fig:12}
\end{figure*}

\par
Figures~\ref{fig:10}--\ref{fig:12} \textcolor{black}{present} a simulated time‑lapse series illustrating the time‑dependent accretion flow and photon ring structure around a rotating Hayward BH, \textcolor{black}{generated with our GRF‑based model}. An animated version of the simulation is available at \url{https://www.youtube.com/@GuoSen-r1c/videos}. The series \textcolor{black}{captures} the evolution of surface brightness fluctuations across the disk, \textcolor{black}{revealing} spiral arms and filamentary structures \textcolor{black}{seeded by} the stochastic, nonstationary properties of the GRF. Each frame \textcolor{black}{represents} a discrete time slice, together \textcolor{black}{portraying} the morphological evolution of the disk. Notably, localized segments of the photon ring and shadow boundary undergo transient brightening, \textcolor{black}{suggestive of} ``hot spot” formation driven by the interplay of spacetime curvature and \textcolor{black}{magnetically induced} perturbations in the rotating Hayward spacetime.

\par
Adopting the \textit{fast‑light} approximation, each frame in the simulation sequence represents a \textcolor{black}{frozen‑time snapshot} of the disk emission, effectively \textcolor{black}{bypassing the need to compute individual photon time‑of‑flight delays}. This approximation retains \textcolor{black}{rapid luminosity fluctuations} in the disk and subtle \textcolor{black}{temporal variations} in both the intensity and morphology of the photon ring. Such behavior \textcolor{black}{manifests} the complex interplay between the BH’s spin, magnetic charge, and the accretion‑flow dynamics. Notably, bright, high‑energy features appearing at different azimuthal angles along the photon ring \textcolor{black}{trace} the orbital motion of dense, magnetized plasma, \textcolor{black}{modulated} by frame dragging and magneto‑gravitational coupling inherent to the rotating Hayward spacetime.

\par
The distinction between the \textit{fast‑light} and \textit{slow‑light} approximations lies in the treatment of photon travel times. In the \textit{fast‑light} approach, disk emission is assumed to occur simultaneously across all spatial locations, \textcolor{black}{thereby} \textcolor{black}{simplifying} the radiative‑transfer calculation. Conversely, the \textit{slow‑light} approach \textcolor{black}{explicitly accounts for} time‑of‑flight delays, computing each photon’s emission and arrival times independently to preserve the \textcolor{black}{temporal causality} of light propagation. While the \textit{fast‑light} approximation is \textcolor{black}{computationally efficient} and captures many \textcolor{black}{qualitative} features, \textit{slow‑light} provides a more \textcolor{black}{physically accurate} depiction of systems exhibiting \textcolor{black}{rapid} time variability or \textcolor{black}{strong} relativistic effects.

\par
The \textcolor{black}{principal} distinction between the \textit{fast-light} and \textit{slow-light} approximations \textcolor{black}{resides} in the treatment of photon propagation \textcolor{black}{delays}. In the \textit{fast-light} \textcolor{black}{approximation}, disk emission is \textcolor{black}{assumed instantaneous} across all spatial locations, thereby neglecting light-travel time effects. \textcolor{black}{This} simplification \textcolor{black}{substantially} reduces computational \textcolor{black}{expense} while preserving the \textcolor{black}{essential} qualitative features of the resulting images. Conversely, the \textit{slow-light} approximation \textcolor{black}{explicitly} accounts for the full photon time-of-flight by computing emission and arrival times along each null geodesic. \textcolor{black}{By retaining the causal sequence of photon propagation}, this approach enables more accurate modeling of \textcolor{black}{systems exhibiting rapid temporal variability, strong relativistic lensing, or asymmetric brightness evolution}. Although \textcolor{black}{considerably} more computationally demanding, slow-light ray tracing \textcolor{black}{affords enhanced} fidelity in capturing the time-dependent morphology of the BH environment.

\section{Conclusion and Discussion}
\label{sec:4}
\par
In this study, we have \textcolor{black}{developed a unified} framework for \textcolor{black}{simulating and} imaging time‑dependent accretion flows around rotating Hayward BHs by \textcolor{black}{generalizing} the Matérn field formalism to the spatio‑temporal domain. \textcolor{black}{By leveraging} a GRF approach, we \textcolor{black}{construct} fluctuation fields characterized by \textcolor{black}{locally} anisotropic and inhomogeneous covariance structures that faithfully \textcolor{black}{reproduce} the statistical signatures of turbulent GRMHD flows. These stochastic fields are then \textcolor{black}{incorporated} into both fast‑light and slow‑light ray‑tracing schemes to \textcolor{black}{generate} high‑fidelity, time‑resolved synthetic images of BH environments.

\par
One of the principal \textcolor{black}{advancements} of this study is the incorporation of \textcolor{black}{spatially inhomogeneous, tensorial correlation structures} via the local anisotropy tensor \(\Lambda(\mathbf{x})\), which \textcolor{black}{encodes} direction‑dependent coherence scales derived from realistic disk geometry and Keplerian velocity fields. Furthermore, we have \textcolor{black}{shown} how temporal evolution can be incorporated into the emission model through a time‑dependent SPDE, producing dynamically coherent features such as spiral arms and transient hotspots across the disk.

\par
By \textcolor{black}{conducting simulations under} both fast‑light and slow‑light \textcolor{black}{observational paradigms}, we \textcolor{black}{quantified} the influence of light propagation delays on image morphology. \textcolor{black}{Specifically}, slow‑light simulations \textcolor{black}{expose} temporally smeared features and more faithful representations of causally delayed emission, \textcolor{black}{thereby enhancing physical realism} in strongly lensed or highly variable source regimes. Our synthetic images are \textcolor{black}{qualitatively} consistent with state‑of‑the‑art GRMHD simulations while \textcolor{black}{achieving substantial computational savings by leveraging} a stochastic generative model.
\begin{figure*}[htbp]
  \centering
  \includegraphics[width=14cm,height=6cm]{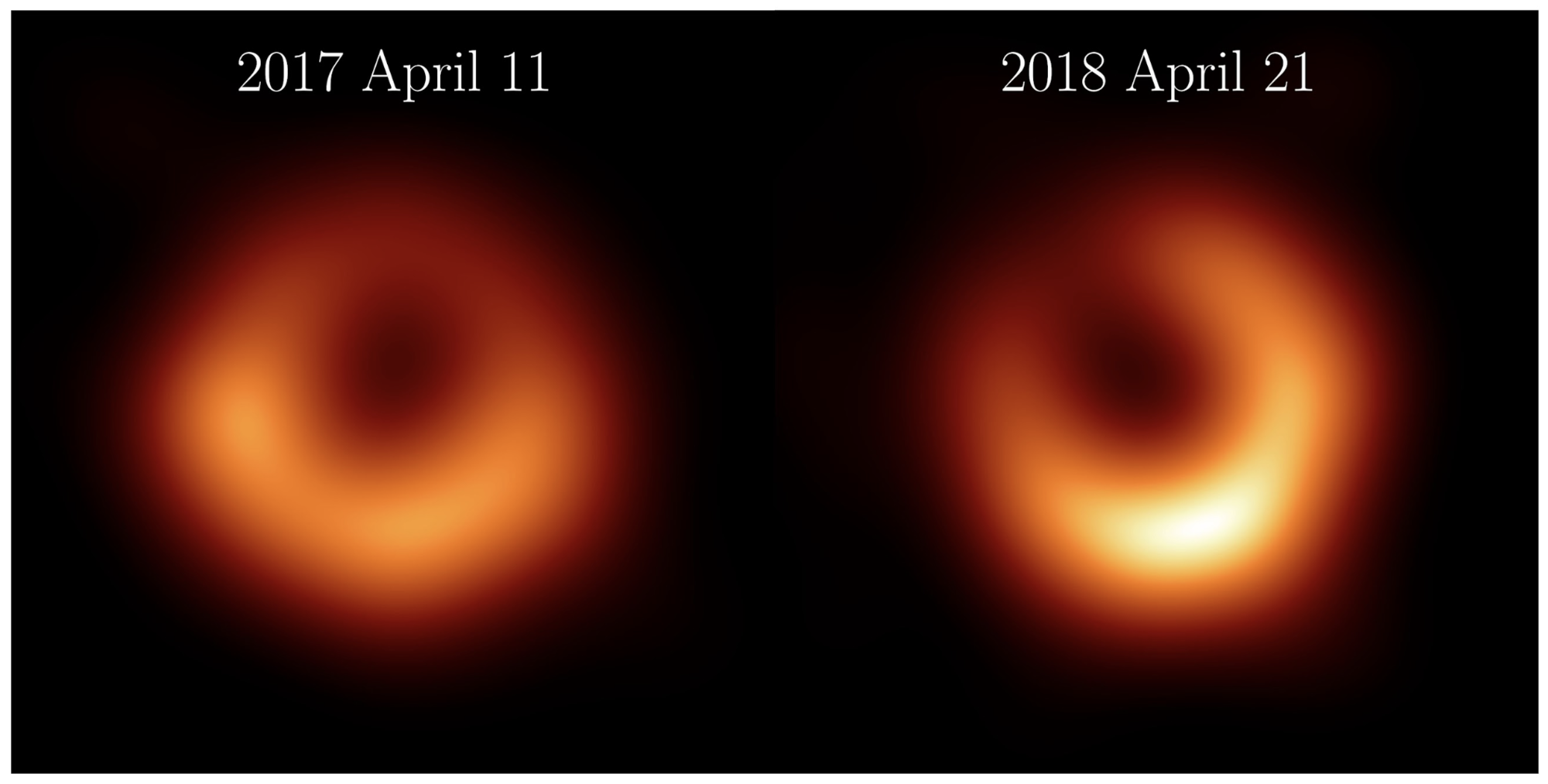}
  \caption {The EHT Collaboration has released new images of M87$^{*}$ from observations taken in April $2018$, one year after the first observations in April $2017$. This bright ring surrounds a dark central shadow, and the brightest part of the ring in $2018$ has shifted by about $30^\circ$ relative from $2017$ to now lie in the $5$ o’clock position~\cite{42}.}\label{fig:13}
\end{figure*}

\par
Our \textcolor{black}{time‑resolved imaging simulations} of a rotating Hayward BH provide \textcolor{black}{new} insights into \textcolor{black}{temporal} features observed in M87$^{*}$. The 2018 EHT images \textcolor{black}{revealed} a \textcolor{black}{systematic} shift in the brightest sector of the emission ring, \textcolor{black}{advancing} clockwise by approximately $30^\circ$ from the 2017 observation to a location near the “5 o’clock” azimuth. This \textcolor{black}{temporal displacement} indicates significant time‑dependent variations in the high‑energy emission regions \textcolor{black}{proximate} to the BH. We \textcolor{black}{further investigate} how the evolution of the accretion disk influences black‑hole imaging. In \textcolor{black}{scenarios} where the BH itself exhibits time‑varying behavior—such as in Vaidya spacetimes—these \textcolor{black}{dynamic} effects can \textcolor{black}{imprint} directly on the observed images. \textcolor{black}{Comprehensive} details of these analyses, including \textcolor{black}{alternative} models and broader astrophysical implications, will be \textcolor{black}{presented} in future work.

\par
\textcolor{black}{In future work}, we \textcolor{black}{plan to extend our model to incorporate polarimetric observables, implement radiative-feedback processes}, and \textcolor{black}{integrate fully general–relativistic radiative-transfer solvers}. The framework \textcolor{black}{developed} here \textcolor{black}{constitutes a versatile and powerful platform} for exploring a broad class of regular BH metrics and \textcolor{black}{can be directly applied to interpret data} from forthcoming high‑resolution interferometric campaigns such as the EHT and ngEHT.

\acknowledgments

This work is supported by the National Natural Science Foundation of China (Grant No. 12133003), Fapesq-PB of Brazil, the Fund Project of Chongqing Normal University (Grant Number: 24XLB033), and the Key Project of Sichuan Science and Technology Education Joint Fund (25LHJJ0097).

% The bibliography will probably be heavily edited during typesetting.
% We'll parse it and, using the arxiv number or the journal data, will
% query inspire, trying to verify the data (this will probalby spot
% eventual typos) and retrive the document DOI and eventual errata.
% We however suggest to always provide author, title and journal data:
% in short all the informations that clearly identify a document.

\end{document}